\documentclass[aps,prb,epsf,epsfig,floatfix,showpacs,groupedaddres,superscriptaddress,footinbib,twocolumn]{revtex4-2}
\usepackage{amsmath, amssymb, txfonts, bm, graphicx, subcaption, float}
\usepackage{anysize}
\usepackage[margin=1in]{geometry}
\usepackage{enumerate}
\usepackage{amsbsy}
\usepackage[font=small]{caption} 
\usepackage{subcaption, float}
\numberwithin{equation}{section}
\usepackage{algorithm,algorithmic}
\usepackage{url}

\usepackage{makecell}
\usepackage{listings}

\usepackage[usenames,dvipsnames,svgnames,table]{xcolor}
\usepackage{array}
\usepackage{tabularray}
\usepackage{colortbl} 
\usepackage{cellspace} 
\usepackage{pstricks}
\definecolor{Blue}{rgb}{0.0,0.0,1.0} 
\definecolor{Green}{rgb}{0.1,1,0.8} 
\definecolor{Red}{rgb}{0.6,0.1,0.1} 
\definecolor{backcolour}{rgb}{0.95,0.95,0.92}
\definecolor{Gray}{rgb}{0.80784, 0.86667, 0.90196} 
\definecolor{Lightgray}{rgb}{0.9176, 0.95, 0.95686} 
\definecolor{Akzent}{rgb}{0.6627, 0.63529, 0.55294} 
\definecolor{armygreen}{rgb}{0.29, 0.33, 0.13} 
\definecolor{darkgreen}{rgb}{0.0, 0.2, 0.13}
\definecolor{tcol}{rgb}{0,0,1} 
\definecolor{tcoll}{HTML}{013220}  
\newrgbcolor{dgreen}{0.0 0.5 0.2}
\newrgbcolor{pink}{1.0 0.0 0.4}
\newrgbcolor{violet}{0.6 0.0 0.4}
\newrgbcolor{orange}{1.0 0.2 0.0}
\newrgbcolor{dred}{0.8 0.1 0.0}
\def\tcol{\color{tcol}}
\def\tcoll{\color{dgreen}}
\usepackage{verbatim}
\usepackage{hyperref}
\hypersetup{
     colorlinks=true,
     citecolor=blue,
     urlcolor=violet,
     linkcolor=red
}
\usepackage{empheq}

\lstdefinestyle{PythonStyle}{
    language=Python,
    basicstyle=\ttfamily\footnotesize,
     backgroundcolor=\color{backcolour},
     keywordstyle=\color{Blue},
     commentstyle=\color{Green},
     stringstyle=\color{Red},
     breaklines=true,
     numbers=left,
    stepnumber=1,
    frame=single,
    showspaces=false,
    showstringspaces=false,
    showtabs=false,
    tabsize=4,
    captionpos=b,
    breaklines=true,
    breakatwhitespace=true
}
\def\FIGDIR{.}
\def\blgn#1\elgn{\begin{align}#1\end{align}}

\def\h{\hat}
\def\al{\alpha}
\def\bt{\beta}
\def\del{\delta}
\def\g{\gamma}

\def\Phip{\phi^+}

\def\obst{\frac{1}{\sqrt{2}} } 
\newcommand{\bbmat}{\begin{bmatrix}}
\newcommand{\ebmat}{\end{bmatrix}}

\usepackage{bbm}
\usepackage{braket}
 
\def\non{\nonumber}

\newcommand{\eref}[1]{Eq. (\ref{#1})}

\newcommand{\fref}[1]{Fig.~\ref{#1}}
\newcommand{\tref}[1]{Table~\ref{#1}}
\newcommand{\sref}[1]{Sec.~\ref{#1}}

\def\h{\hat}
\def\hf{\frac{1}{2}}

\def\bi{\begin{itemize}}
\def\ei{\end{itemize}}
\def\i{\item}
\def\ketz{\ket{0}}
\def\keto{\ket{1}}
\def\ketq{\ket{Q}}
\def\brao{\bra{1}}
\def\braz{\bra{0}}
\def\Phip{\ket{\Phi^+}}
\def\Phim{\ket{\Phi^-}}
\def\Psip{\ket{\Psi^+}}
\def\Psim{\ket{\Psi^-}}
\def\ketzz{\ket{00}}
\def\brazz{\bra{00}}
\def\ketzo{\ket{01}}
\def\brazo{\bra{01}}
\def\ketoz{\ket{10}}
\def\braoz{\bra{10}}
\def\ketoo{\ket{11}}
\def\braoo{\bra{11}}
\newcommand{\bnu}{\begin{enumerate}}
\newcommand{\enu}{\end{enumerate}}

\def\etal{\emph{et al.}}
\def\ss{\subsection}
\def\sss{\subsubsection}
\begin{document}

\title{Quantum Teleportation Game
- A fun way to play and learn single qubit teleportation protocol
}
\author{Himadri Barman}\email{reducedpc@gmail.com}
\affiliation{Department of Physics, Zhejiang University, Hangzhou 310027, China}
\begin{abstract}
  We demonstrate how the quantum teleportation protocol of a single qubit can be understood
by designing a simple game that can be played by three participants: Alice, Bob, and 
\emph{Quantum God}. 
\end{abstract}
\maketitle

\section{What is quantum teleportation?}
\begin{figure}[!hp]
    \includegraphics[height=4cm,clip]{\FIGDIR/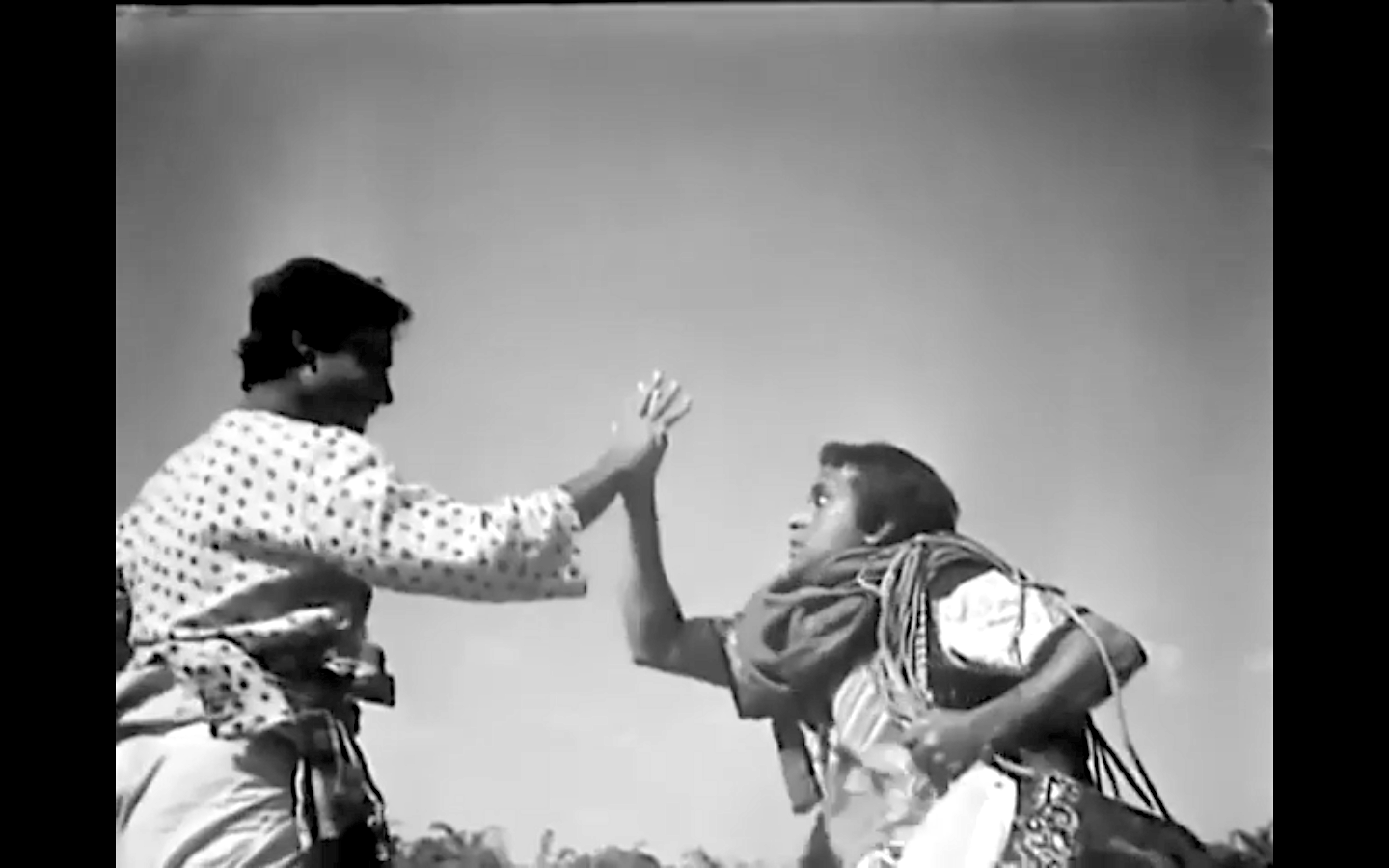}
\caption{
A scene of teleportation from the iconic film Goopy Gyne Bagha Byne (1969), directed by Academy Award-winning filmmaker Satyajit Ray. The protagonists, Goopy Gyne and Bagha Byne, successfully teleport themselves by joining hands and shouting the name of their desired destination.}
\label{fig:ggbb:teleport}
\end{figure}

  Teleportation refers to sending an object fast to a desired location without following a conventional trajectory-based path.
This phenomenon has been depicted in several science fiction and fantasy stories. To mention, Indian filmmaker Satyajit Ray's
iconic film Goopy Gyne Bagha Byne (1969)'s~\cite{ray:ggbb:69} duo protagonists had the magical power to teleport themselves by a snap
of a clapping together (see ~\fref{fig:ggbb:teleport}). Also, who can forget the popular phrase \emph{beam me up} (asking someone to teleport through a \emph{transporter}) in Star Trek's episodes?
Though the above fictional examples deal with transporting real objects (even human beings), \emph{quantum teleportation} (QT) in principle transfers a quantum information (defined by the quantum state) almost instantly from one place to another. 
The simplest and most popular protocol for teleporting a single qubit was first proposed by Bennett \etal~\cite{bennett:etal:prl93}.
%
A few years later, their protocol was experimentally confirmed using photonic qubits~\cite{bouwmeester:etal:zeilinger:gr:nat97,braunstein:kimble:prl98}. Since then, QT has been demonstrated through various platforms, including NMR~\cite{nielsen:laflamme:nat98}, coherent optical modes~\cite{furusawa:etal:sc98,takei:etal:pra05}, trapped ions~\cite{riebe:etal:nat04,barrett:etal:nat04,wan:etal:sc19}, combined light and matter~\cite{sherson:etal:nat06}, solid-sattes~\cite{pfaff:etal:sc14, reindl:etal:sciadv18}, superconducting circuits~\cite{steffen:etal:nat13}, and many other realizations~\cite{pirandola:nphoton15, hu:etal:nrev23}. The QT protocol has also been extended to multiqubits~\cite{boumeester:etal:zeilinger:gr:prl99,pan:etal:zeilinger:gr:prl01,zhao:etal:nat04} and high-dimensional qubits~\cite{mafu:etal:pra13, luo:etal:prl19, zhang:etal:pra19, erhard:krenn:zeilinger:nrp20, hu:etal:prl20}. 
Notably, successful teleportation has been achieved over a record distance of 1,400 km~\cite{ren:etal:nat17}.
QT has become a cornerstone of modern quantum computation and quantum information giving rise to numerous research areas such as quantum key distribution (QKD)~\cite{bennett:brassard:proceed84,ekert:prl91}, quantum internet~\cite{cirac:zoller:kimble:mabuchi:prl97, kimble:nat08, wehner:elkouss:hanson:sc18}, measurement-based computing~\cite{raussendorf:briegel:prl01, briegel:raussendorf:nest:nphys09}, and quantum repeaters~\cite{briegel:dur:cirac:zoller:prl98}. Therefore an understanding of the basic protocol is is highly desirable. If this understanding can be achieved in an engaging way, such as by playing a simple rules-based game, with fun and entertainment, it may attract greater interest and awareness among students and the general public.
With this spirit in mind, we have designed a quantum teleportation game, which serves as the central focus of our paper. In the forthcoming sections, we first discuss the theory of the single-qubit QT protocol originally proposed by Benett \etal~\cite{bennett:etal:prl93}, then provide a detailed explanation of the game, and finally conclude with a summary.
%
%
%
\section{The teleportation protocol}
\label{sec:qt:protocol}
\begin{figure}[!htp]
\includegraphics[height=4cm,clip]{\FIGDIR/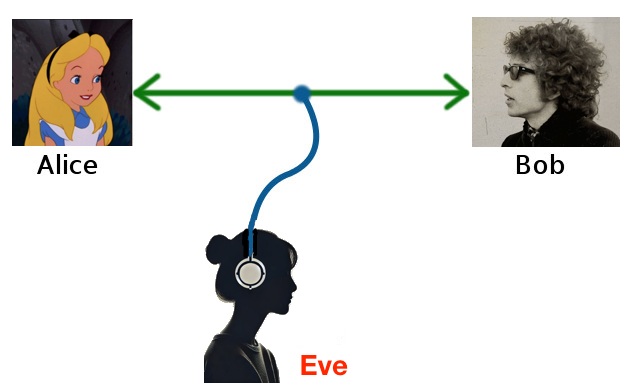}
\caption{A schematic of quantum teleportation: Alice sends quantum information to Bob with whom she shares
an entangled state. Meanwhile, Eve tries to intercept the information but fails as the entanglement between Alice and Bob keeps the information protected.}
\label{fig:qt:alice:bob:eve}
\end{figure}

In the QT protocol, an agent named Alice is assigned to send a quantum information (represented by state $\ket{Q}$)
to another agent, Bob, without it being intercepted by any rival or competing agent, Eve (see \fref{fig:qt:alice:bob:eve}). To acheive her objective, Alice prepares an entangled state (a Bell state) shared with Bob. This presence of quantum entanglement allows Bob to generate exactly the same state in his own place after performing some quantum operations.        

We briefly describe the protocol for a single qubit $\ket{Q}$ below.

\subsection*{Initialization and entanglement preparation}
There are three qubits. Two of them are distributed to Alice and the third one is given to Bob. 
The first qubit is in the state $\ket{Q}$ which can be generically written in terms of single qubit bases
$\ket{0}$ and $\ket{1}$ (i.e. superposition of $\ket{0}$ and $\ket{1}$):
\blgn
\ket{Q} = a\ket{0} + b\ket{1}\,
\elgn
where $|a|^2 + |b|^2 = 1$ satisfying the quantum probability conservation.
Alice's task is to deliver the state $\ket{Q}$ to Bob through teleportation 
even though she may not have any specific knowledge about $\ket{Q}$.

\subsubsection*{Convention of denoting multiqubit states:}
\begin{figure}[!htp]
    \includegraphics[height=3cm,clip]{\FIGDIR/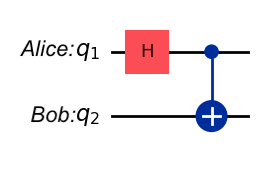}
\caption{Generation of the Bell state $\Phip$ from two qubits both initialized at state $\ketz$: A Hadamard gate operates on the first qubit. A controlled $NOT$ ($CNOT$) or $CX$ gate operates between first and second qubit. Image is adapted from the output generated by Python coding with IBM's \emph{Qiskit} module.}
\label{fig:Phip:state:genrn}
\end{figure}

We adopt the convention of denoting a multiqubit tensor product state by sequencing individual qubit states from right to left. In our 3-qubit situation, if the first, second, and third qubits are $q_0$, $q_1$, and $q_2$ respectively, the convention instructs us to write the 3-qubit state as 
${\tcol \ket{q_2}\ket{q_1}\ket{q_0}=\ket{q_2 q_1}\ket{q_0} = \ket{q_2} \ket{q_1 q_0} = \ket{q_2 q_1 q_0}}$.
Thus, the first and last qubits are denoted on the extreme right and extreme left, respectively. 
Now, let as assume the other qubits of Alice and Bob are initially in states $\ket{A}$ and $\ket{B}$. Then the 
raw 3-qubit state is $\ket\psi_0 = \ket{B A Q}$.
Through a set of quantum operations (will be detailed soon), an entangled state is created and shared between them. 
A common such entanglement state is one of the EPR states or Bell states (conventionally denoted
by $\Phip$, $\Psip$, $\Phim$, and $\Psim$)~\cite{book:mike:ike:cambridge10}. In our case, we choose $\Phip \equiv (\ket{00} + \ket{11})/\sqrt{2}$ as the entanglement 
pair. If both $\ket{A}$ and $\ket{B}$ are in the state $\ket{0}$, then $\Phip$ can be easily generated 
by the following two-step quatum gate operations (see \fref{fig:Phip:state:genrn} for the circuit representation).
\begin{enumerate}[\bfseries{Step}~1:]
\setcounter{enumi}{-1}
\i Operate with a Hadamard ($H$) gate on $\ket{A}=\ket{0}$ and create a superposition state: 
$\ket{A'}=\obst\big[\ketz + \keto]$. This updates 2-qubit state between Alice and Bob to
$\ket{\phi_1} = \ket{B A'}$ and the overall 3-qubit state becomes
\blgn
\ket{\psi_0} \to \ket{\psi_1} 
= \ket{BA'}\ket{Q} = \ket{\phi_1}\ket{Q}\,. 
\elgn
\i Operate a controlled $NOT$ or $CX$ gate where $\ket{A'}$ is the control qubit and $\ket{B}$ is 
the target bit. This modifies $\ket{\phi_1}$ into the Bell state $\Phip$. The state can no longer be expressed as a product of two individual quantum states as it is already entangled.
\end{enumerate}
%
Thus, after setting up an entanglement between Alice and Bob, the ready-to-teleport 
3-qubit state is prepared as  
\blgn
  \ket{\psi_2} &= \Phip\ket{Q}=\obst[\ket{00} + \ket{11}][a\ket{0} + b\ket{1}]\non\\
              &= \obst\big[a\ket{000} + a\ket{110} + b\ket{001} + b\ket{111}\big]\,.
\elgn

\subsection*{Alice's operations}
Now, Alice operates a $CNOT$ or $CX$  gate (denoted by the operator ${\h X}_C$) between the state $\ket{Q}$ ($q_0$) and her own qubit ($q_1$). This updates the state $\ket{\psi_2}$ to $\ket{\psi_3}$: 
\blgn 
 \ket{\psi_3} = {\h X}_C(0,1) \ket{\psi_2} 
&=  \obst\big[a\ket{000} + a\ket{110} \non\\
&\qquad+ b\ket{011} + b\ket{101} 
  \big] \,.
\elgn
\begin{widetext}
Note that the control and target qubits are denoted by the indices of quantum wires 
in the circuit (here $0$ and $1$ for wires $q_0$ and $q_1$) and expressed within the parentheses following a controlled gate operator (here ${\h X}_C$, see \sref{app:basic:qgates}).
\begin{center}
\begin{figure}[!hbp]
    \includegraphics[height=4cm,clip]{\FIGDIR/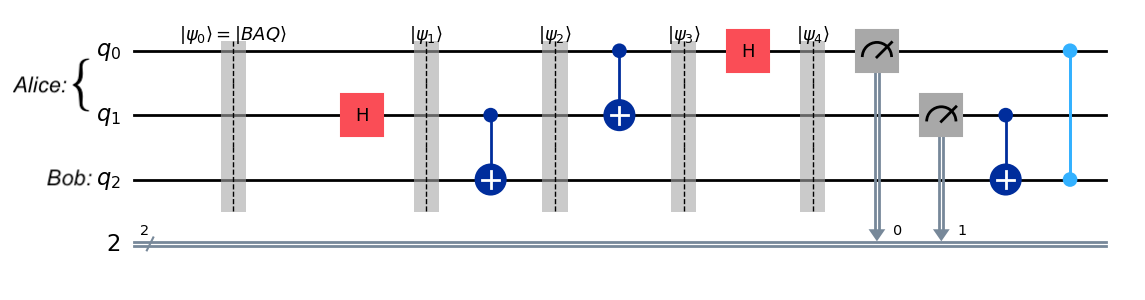}
    \caption{The single qubit quantum state (defined by $\ket{Q}$) teleportation circuit with all necessary gates and measurements. Image is adapted from the output generated by Python coding with IBM's \emph{Qiskit} module.}
\label{fig:qt:full:ckt}
\end{figure}
\end{center}
\end{widetext}
After this, she applies a Hadamard ($H$) gate on $q_0$, which modifies  $\psi_3$ to  $\psi_4$:  
\blgn \ket{\psi_4} &= H(0) \ket{\psi_3} \non\\
                   &= \obst\obst\bigg[ a\ket{00}\big[\ket{0} + \ket{1}\big] 
                         +a\ket{11}\big[\ket{0} + \ket{1}\big]\bigg]\non\\  
                        &\quad+ \obst\obst\bigg[ b\ket{01}\big[\ket{0} - \ket{1}\big]
                                +  b\ket{10}\big[\ket{0} - \ket{1}\big]\bigg] \non\\
&= \hf\bigg[ a\ket{000} + a\ket{001} + a\ket{110} + a\ket{111}\non\\ 
  &\quad+ b\ket{010} -b\ket{011} + b\ket{100} - b\ket{101}\bigg]\non
\elgn 
$\ket{\psi_4}$ can be rearranged as
\blgn
\ket{\psi_4}
&= \hf\bigg[ 
             \big[a\ket{000} + b\ket{100}\big] 
           + \big[a\ket{011} - b\ket{111}\big]\non\\
    &\qquad   
           + \big[a\ket{110} + b\ket{010}\big] 
           + \big[a\ket{111} - b\ket{011}\big]
   \bigg]\non\\
&=\hf\bigg[ 
           \big[a\ket{0}+b\ket{1}\big]\ket{00} 
         + \big[a\ket{0}-b\ket{1}\big]\ket{10}\non\\
     &\qquad
       +   \big[a\ket{1} + b\ket{0} \big]\ket{01} 
       +   \big[a\ket{1} - b\ket{0} \big]\ket{11}
   \bigg]\non\\
& = \hf\bigg[ 
              {\blue\big[{\h 1}\ket{Q}\big]}       \ket{00} 
            + {\blue\big[{\h X}\ket{Q}\big]}       \ket{10}\non\\
      &\qquad
            + {\blue\big[{\h Z}\ket{Q}\big]}       \ket{01} 
            + {\blue\big[{\h X}{\h Z}\ket{Q}\big]} \ket{11}
    \bigg]\,.
\label{eq:qt:protocol:final}
\elgn
Finally, Alice measures the first two qubits available in her lab and reports the outcome 
(through telephone or texting) as two classical bits to Bob.

\subsection{Bob's action}
Bob receives the classical pair of bits from Alice. There are 4 possible outcomes: (i) 00, (ii) 01, (iii) 10, and (iv) 11. Correspondingly, Bob has 
states $\ket{Q}$, ${\h X}\ket{Q}$, ${\h Z}\ket{Q}$, and ${\h X}{\h Z}\ket{Q}$ (see \eref{eq:qt:protocol:final}). For (i), Bob already has created the 
state $\ket{Q}$. So he does not have to do anything further or in other words, he applies an identity operator $\h I$ ($I$ gate) on his present state. 
For other outcomes, Bob needs to operate with the following gates on his existing quantum state. Bob needs to operate with 
an $X$ gate for (ii), a $Z$ gate for (iii), and consecutively $X$ and $Z$ gates for (iv) 
as these operations yield the state $\ket{Q}$. [Note that to get $\ket{Q}$ from $\h{X}\h{Z}\ket{Q}$, we need to operate the latter state with $(\h{X}\h{Z})^\dag = \h{Z}\h{X}$, which means an $X$ gate has to be operated on the state first and then a $Z$ gate will be operated on that.]    
This leads to a lookup table for Bob which he can blindly follow as instructions to generate the unknown state $\ket{Q}$ in his 
own lab:
\\  
%
\definecolor{TopRow}{rgb}{0.4,0.7,1}
\definecolor{NormalRow}{rgb}{0.8,0.9,1}
\begin{table}[!h]
\centering
    \begin{tblr}{
     colspec = {|c|c|c|c|},
     row{1}  = {bg=TopRow, fg=black, j}, 
     row{2}  = {bg=NormalRow, fg=black, c}, 
     row{3}  = {bg=NormalRow, fg=black, c}, 
     row{4}  = {bg=NormalRow, fg=black, c},
     row{5}  = {bg=NormalRow, fg=black, c}
     } 
    \hline
    {\bf Case} &{\bf Alice's\\ classical bits} &{\bf Bob's operators\\ to create \\$\ket{q_2}=\ket{Q}$} &{\bf Compact \\formula}\\
    \hline
     1&{00}                    &${\h I}$  &${\h Z}^0{\h X}^0$\\      
       \hline
     2&{01}                    &${\h Z}$   &${\h Z}^1{\h X}^0$\\
     \hline
     3&{10}                    &${\h X}$   &${\h Z}^0{\h X}^1$\\
     \hline
     4&{11}                    &${\h Z \h X}$   &${\h Z}^1{\h X}^1$\\
     \hline
   \end{tblr}
\caption{A table for Bob telling him to apply the necessary gate(s) to generate $\ket{Q}$ for his qubit $q_2$ 
for the four possible outcomes of Alice's measurement.}
\label{tab:bob:table}
\end{table}
\\
From the table above (\tref{tab:bob:table}), we also notice two things: ${\h X}$ gate only operates when the second qubit ($q_1$) is 
in state $\keto$ (Case 3 and 4) while ${\h Z}$ gate only operates when the first qubit ($q_1$) is 
in state $\keto$ (Case 2 and 4). Thus, Bob merely needs to first apply $CNOT(1,2)$ or $CX(1,2)$ gate and then $CZ(0,2)$ gate on the existing 
3-qubit state. In this way, Bob successfully generates the quantum information $\ketq$ in his place while Alice 
loses it and QT happens (see \fref{fig:qt:full:ckt} for full quantum circuit diagram). Destruction of $\ketq$ in Alice's lab respects the \emph{No-Cloning Theorem}~\cite{wootters:zurek:nat82}, 
i.e. no quantum states can be copied without changing the original states.    

\section{The teleportation game} 
\ss{Preparation} 
There are three players in the game, namely Alice, Bob, and Quantum God (QG). Their roles are the following.
\bi
\i {\bf QG:} QG is responsible for all quantum rules governing in the world. They also help in telling the measurement outcome.
\i {\bf Alice:} Alice has her own qubit $\ket{A} = \ket{0}$ at the beginning. But she acquires another qubit $\ket{Q}$  from QG, 
which she has been instructed to teleport to her friend Bob. 
\i {\bf Bob:} Bob is supposed to receive the quantum information from Alice and generate the state $\ket{Q}$ in his own lab by performing some quantum gate operations. He gets a default qubit $\ket{B} = \ket{0}$. 
\ei
%
%
%
\begin{figure}[!hbp]
  \centering
     \begin{subfigure}[b]{0.3\textwidth}
         \centering
    \includegraphics[height=3cm,clip]{\FIGDIR/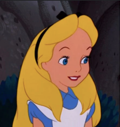}
    \caption{}
    \end{subfigure}
    \begin{subfigure}[b]{0.3\textwidth}
          \includegraphics[height=3cm,clip]{\FIGDIR/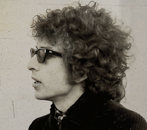}
    \caption{}
    \end{subfigure}
    \begin{subfigure}[b]{0.3\textwidth}
          \includegraphics[height=3.5cm,clip]{\FIGDIR/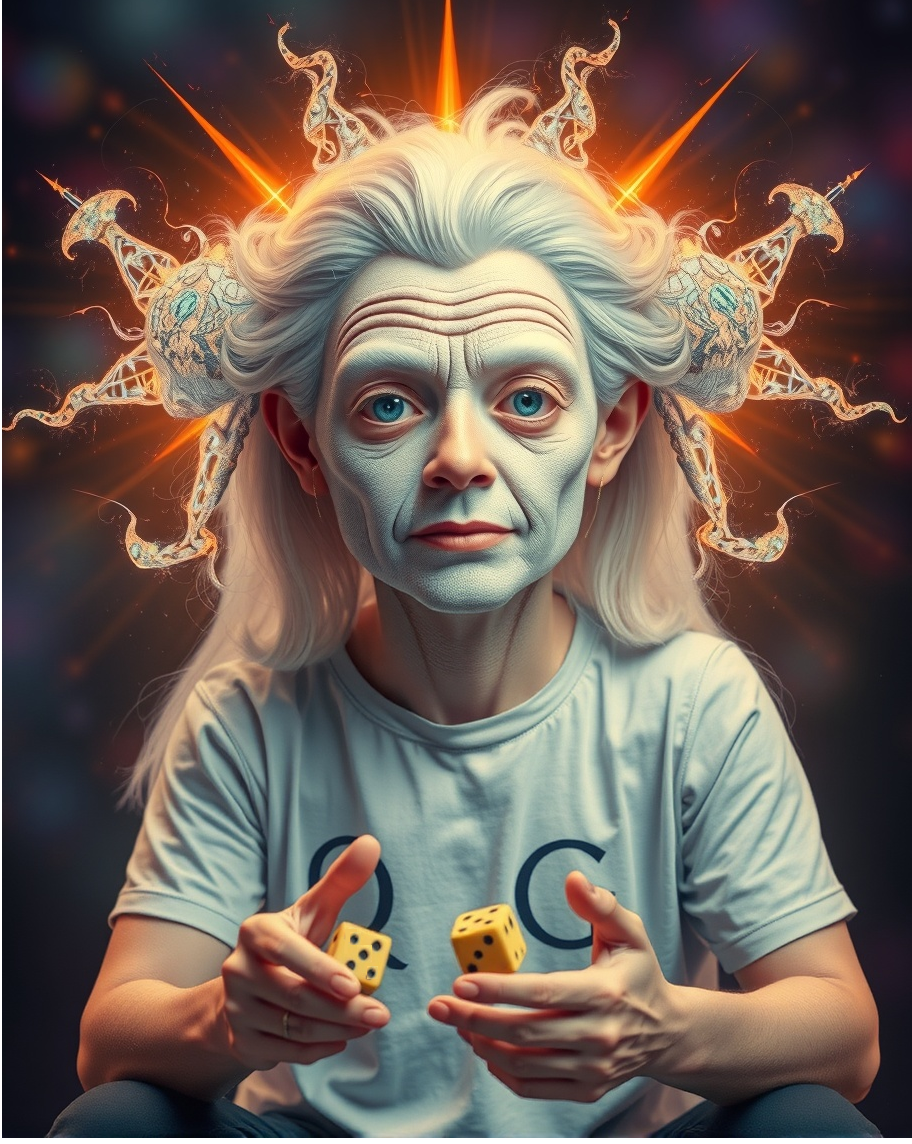}
    \caption{}
    \end{subfigure}
\caption{Players of the Quantum Teleportation Game: (a) Alice, (b) Bob, and (c) Quantum God (representative images). Quantum God (QG) supervises all the quantum rules and updates the quantum states during the running of the game.}
\label{fig:alice:bob:qg}
\end{figure}
\ss{Acting like quantum}
Since the game is only for a demonstration purpose and real quantum equipment are not available, we adopt some classical actions conducted by the players or performers in the game. We furnish below three important quantum parts that can be classically enacted in the game. 

\sss{Updation of a quantum state}
First of all, any quantum information or state is updated by the QG. QG keeps a register (can be a physical notebook or diary), where initial state is noted down by them and evolution of the state after a gate operation is updated by them on the same register. They do not share the quantum information with Alice and Bob at all (like any real quantum state is unknown to an observer). 

\begin{figure}[!hbp]
  \centering
   \begin{subfigure}[b]{0.5\textwidth}
      \centering
       \includegraphics[height=3cm,clip]{\FIGDIR/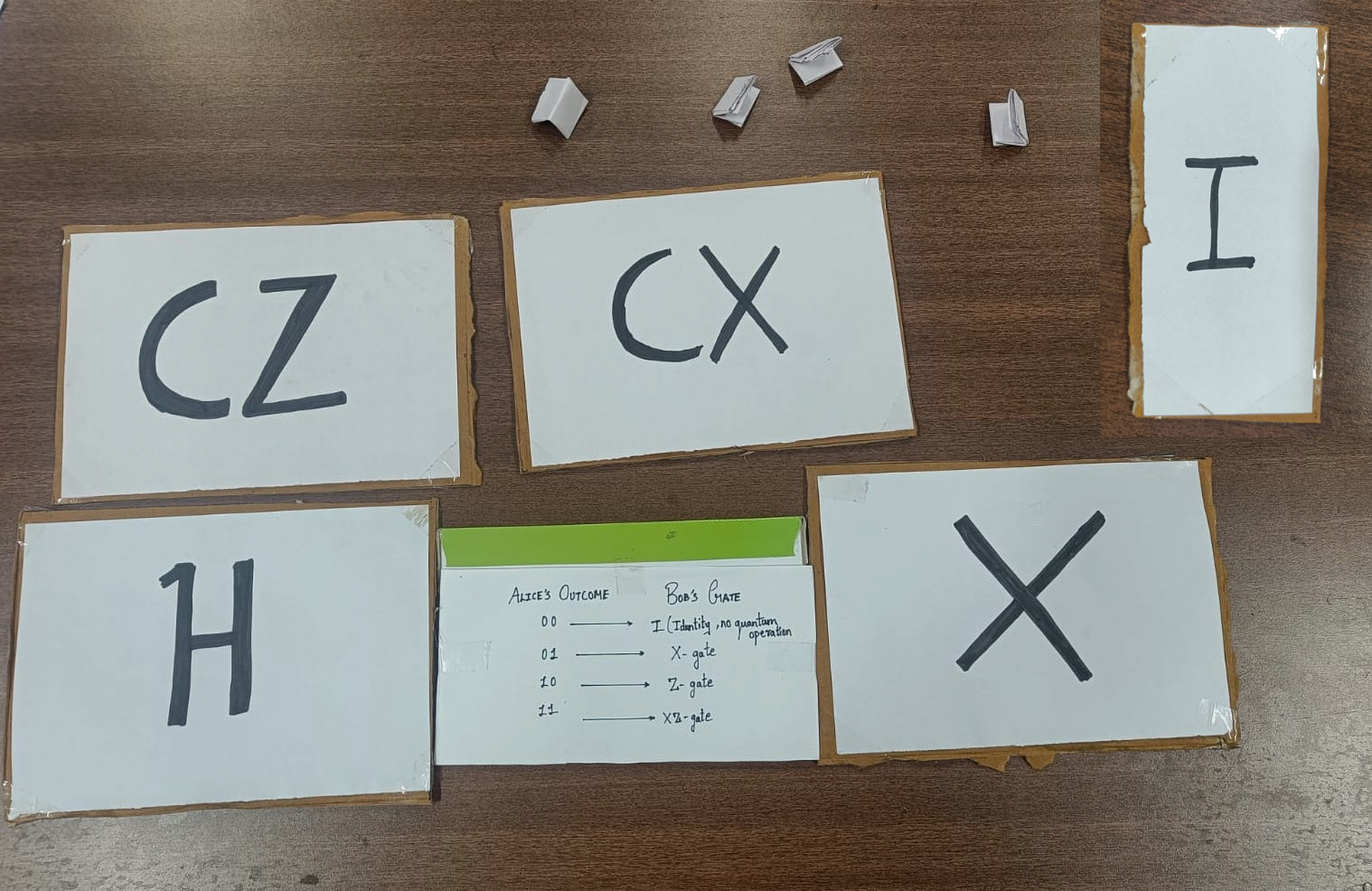}
      \caption{}
    \label{fig:equips:cardboards}  
    \end{subfigure}
   \begin{subfigure}[b]{0.2\textwidth}
      \centering
      \includegraphics[height=3cm,clip]{\FIGDIR/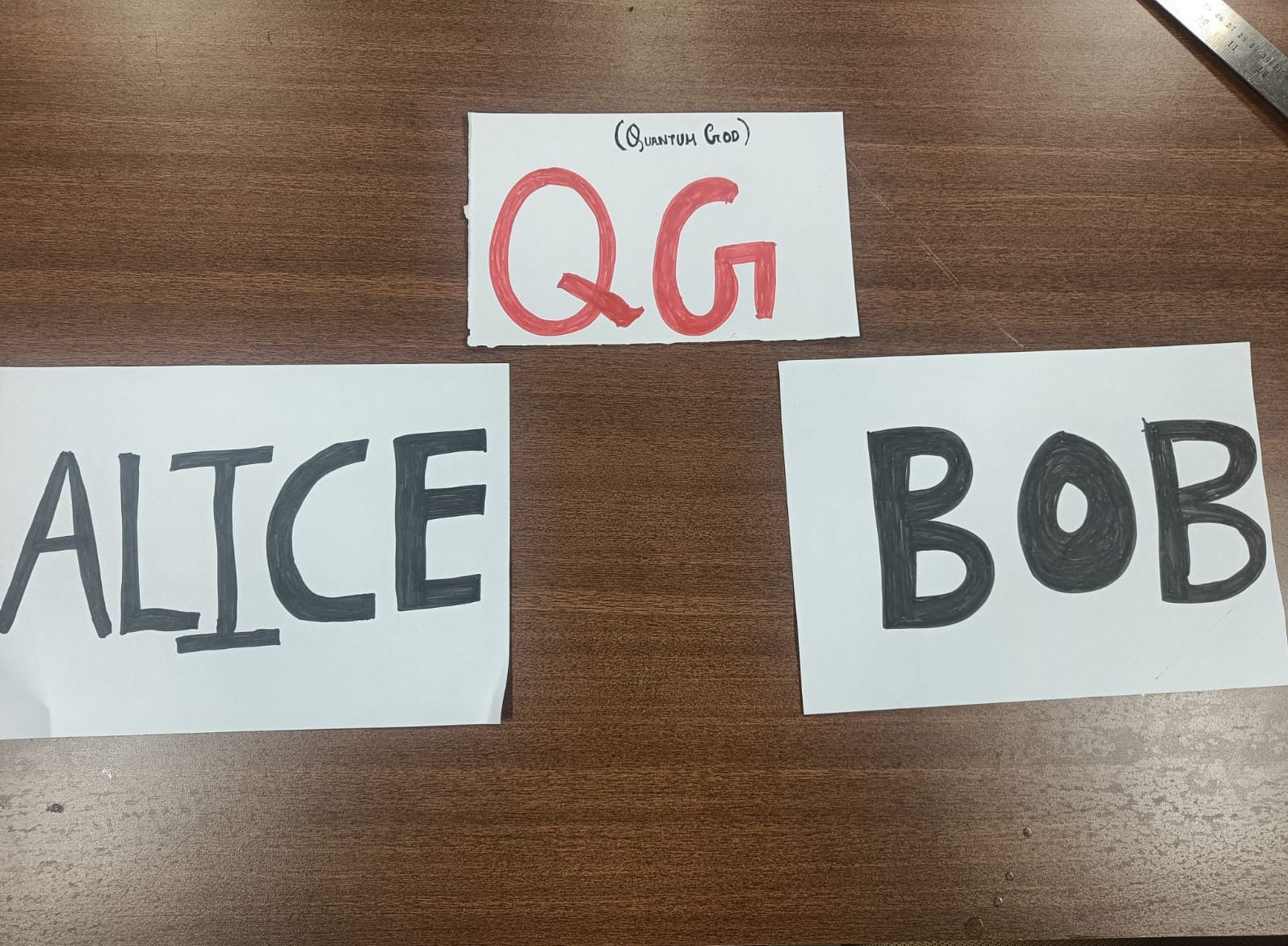}
      \caption{}
      \label{fig:equips:namelabels}  
    \end{subfigure}
        \hspace*{0.5cm}\begin{subfigure}[b]{0.2\textwidth}
      \centering
       \includegraphics[height=3cm,clip]{\FIGDIR/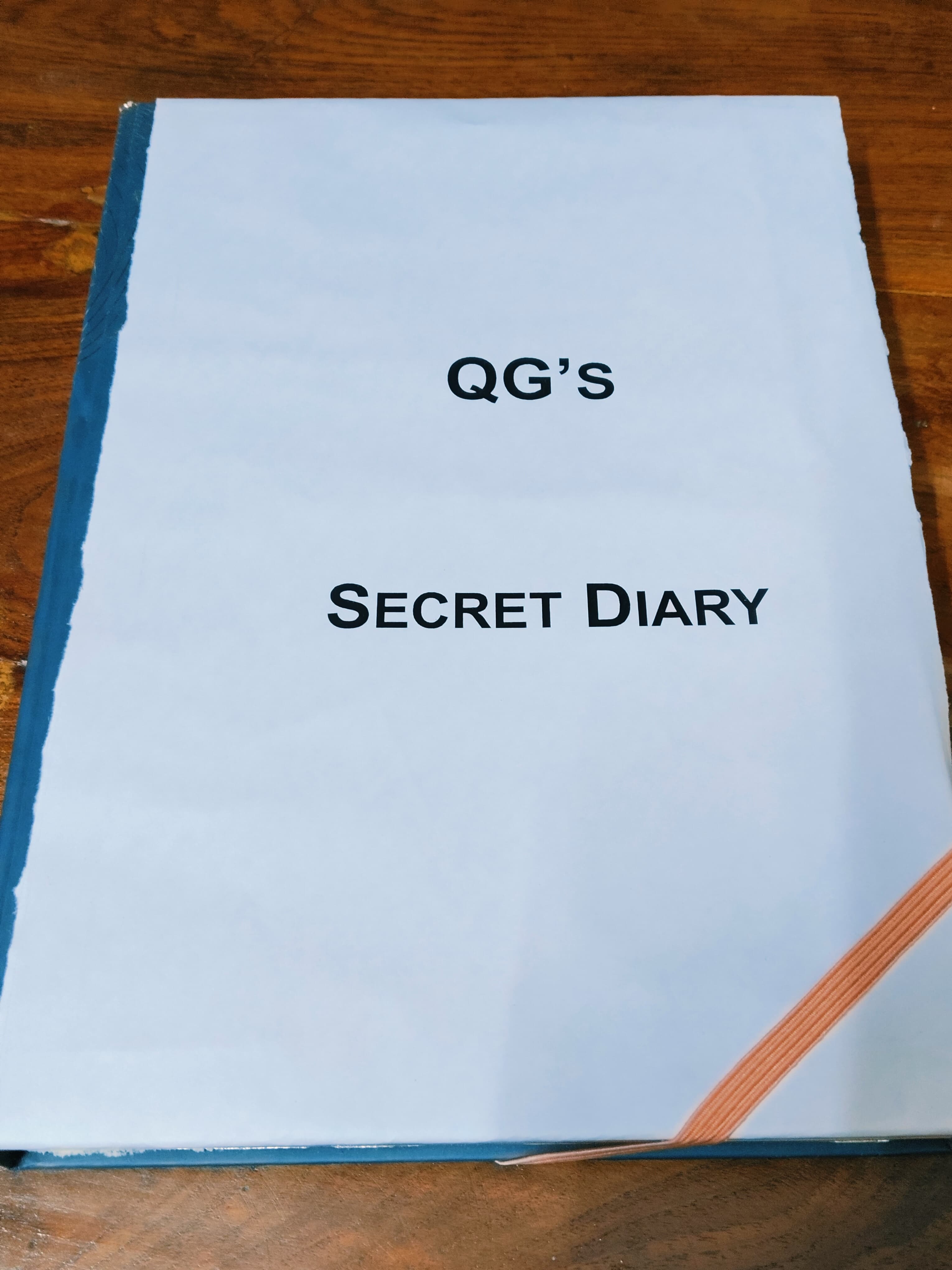}
      \caption{}
    \label{fig:equips:diary}  
    \end{subfigure}
\caption{The basic equipment for the quantum teleportation game: (a) labels to be attached to the players; (b) Cardboard representing gate operators, quantum chits, and Bob's chart; (c) QG's secret diary.} 
\label{fig:equips}  
\end{figure}

\sss{Gate operations}
When Alice or Bob operates a gate, she or he picks up a cardboard where the name of the gate is written and submits it to QG (see~\fref{fig:equips}). QG updates the state after the gate operation as mentioned above. 

\sss{Quantum measurement}
When Alice or Bob makes a quantum measurement, a superposition quantum state gets collapsed to one of the basis state (a superposition state is a linear combination of all the states). 
QG provides a paper chit which  
We create a few placards or cardboards reading the names of the game. When Alice or Bob operates a gate, she or he 
\ss{Equipment}
Following the above discussion, the minimal required pieces of equipment are listed below. 
\bnu
\i Cardboard pieces or cards with gate names: $CX$, $H$, $CZ$, $X$.
\i Nameplates or mugshot boards that can be worn by ``Alice'', ``Bob'', and ``QG''.
\i 4 small pieces of papers where $00$, $01$, $10$, and $11$ will be written.
\i The secret diary of QG.
\enu

\ss{The game}
For the demonstration purpose, let us consider $\ket{Q} = \keto$. Alice and Bob both have state
$\ketz$. Now, the game is played by performing the following actions.
\begin{enumerate}[\bfseries{Action}~1:]
\setcounter{enumi}{-1}
\i QG notes down the initial 3-qubit state on their diary:
 \blgn
    {\tcol \ket{\psi_0}} = \ket{0}\ket{1}\ket{1} = \ket{001}
  \elgn
%
\i  Alice performs an $H$-gate operation. She submits an $H$-gate card to QG (see \fref{fig:alice:submits:gate:to:qg}). This converts her qubit 
to a superposition state $\obst\big[\ketz + \keto\big]$.
QG accepts the gate and updates the state in the diary: 
(see \fref{fig:alice:submits:gate:to:qg}):
  \blgn 
  {\tcol \ket{\psi_1}} 
  &= {\h H}(1)\ket{\psi_1} 
  = \ket{0}\obst\big[\ketz + \keto\big]\ket{1}\non\\ 
  &= \obst\big[\ket{001} + \ket{011}\big]\,.  
  \elgn
\begin{figure}[!hbp]
  \centering
   \begin{subfigure}[b]{0.4 \textwidth}
      \centering
      \includegraphics[height=4.5cm]{\FIGDIR/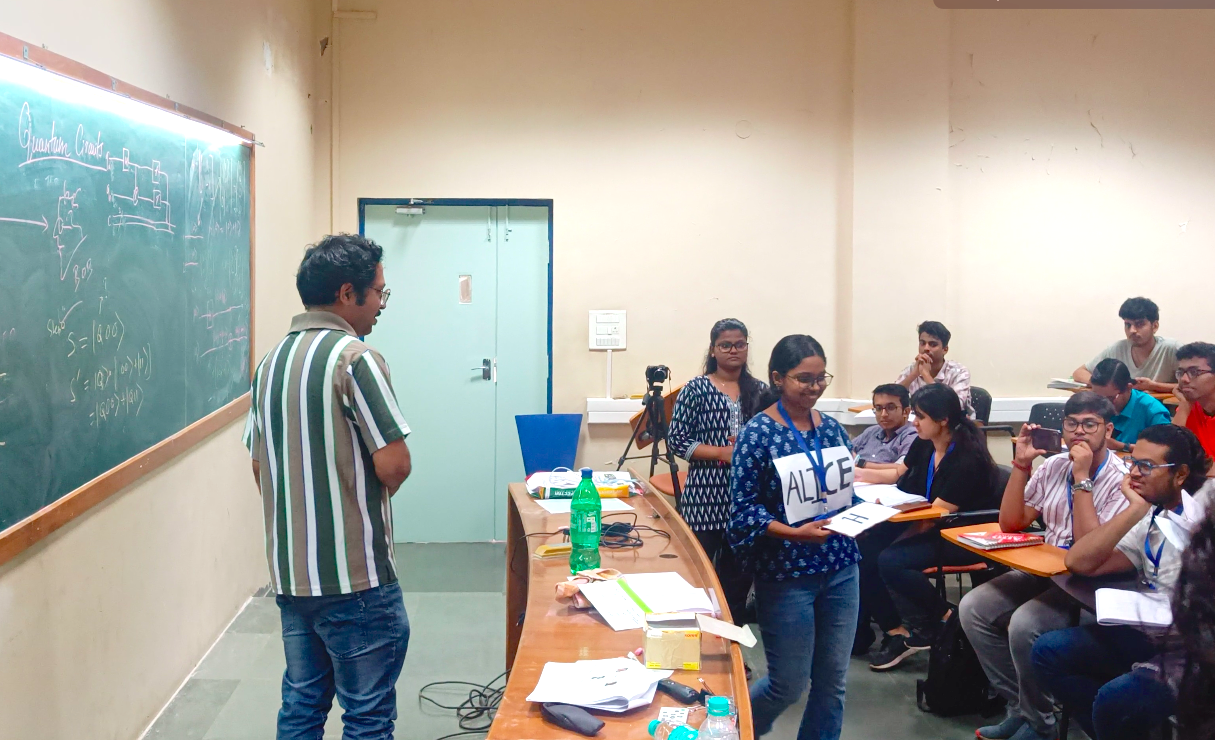}
      \caption{}
      \label{fig:alice:submits:hgate}  
    \end{subfigure}
    \begin{subfigure}[b]{0.4\textwidth}
      \centering
       \includegraphics[height=4.5cm]{\FIGDIR/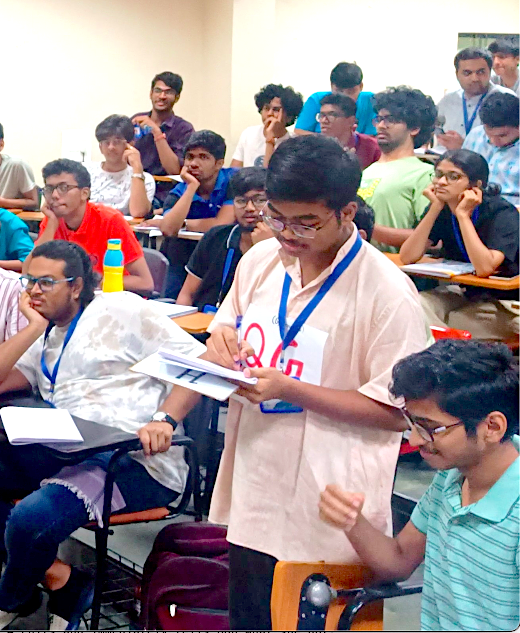}
      \caption{}
    \label{fig:qg:accepts:hgate}  
    \end{subfigure}
\caption{(a) Alice submitting $H$ gate to QG. (b) QG accepting the gate and updating the state on their diary. Images are from the event 
performed at the NIUS Physics Camp 2024, HBCSE, Mumbai, India.} 
\label{fig:alice:submits:gate:to:qg}  
\end{figure}
%
\i   
Now, Alice performs a $CNOT$ or $CX$ operation on (1,2) ($q_1$: controlled bit, $q_2$ target bit). She hands over a 
$CX$ card to QG and mentions the qubits where the gate to be operated.
QG looks into the state and flips the qubit $q_2$ only when the $q_1$
qubit in found to be $\keto$.
Thus, QG updates in the diary:  
 \blgn 
   {\tcol \ket{\psi_2}} = {\h X}_C(1,2) \ket{\psi_1} =  \obst\big[\ket{001} + \ket{111}\big]\,.  
 \elgn
Note that this state is nothing but $\Phip\ket{Q}$ and hence successfully an entanglement is established, through a Bell pair $\Phip$, 
between Alice and Bob's qubits.
%
\i
Alice submits another $CX$ card to QG but now she tells that she wants to operate on (0,1). QG
flips $q_1$'s state only when $q_0$ is found to be in $\keto$. Hence, QG updates:  
\blgn 
   {\tcol \ket{\psi_3}} = {\h X}_C(0,1) \ket{\psi_2}  = \obst\big[\ket{011} + \ket{101}\big]\,.
\elgn
\i
  Alice performs an $H$-gate operation on $q_0$.
  As soon as as she submits the gate's card to QG, they update the state
   (see \fref{fig:alice:submits:gate:to:qg}): 
    \blgn
     {\tcol\ket{\psi_4}} =  {\h H}(0) \ket{\psi_3} 
      &= \hf\ket{01}\big[\ket{0} - \ket{1} \big]
            + \hf\ket{10}\big[\ket{0} - \ket{1} \big]\non\\ 
      &= \hf\big[\ket{010} - \ket{011} + \ket{100} - \ket{101}\big]\,. 
     \elgn       
\i
 Alice makes measurements of the first and second qubits.  
 QG provides her with 4 chits that contain the 4 possible outcomes:
 01, 11, 00, and 10.  
 Alice plays a lottery game -- she randomly picks one out of the 4 chits. This action mimics   
quantum measurement which is probabilistic. She immediately tells Bob what she got. In the game, Alice hands over her chit to Bob ((see \fref{fig:alices:chit:bobs:chart}). 
%
\i 
Bob reads the pair of binary digits from the chit and tallies them with a chart.
We call it \emph{Bob's chart}, derived from \tref{tab:bob:table}, which simply tells
what gates he needs to operate on his present quantum state in order to convert his own 
qubit $q_2$'s state into $\ket{Q}$ (see \fref{fig:alices:chit:bobs:chart:c}). 
The chart reads
\begin{table}[!h]
\centering
    \begin{tblr}{
     colspec = {|c|c|},
     row{1}  = {bg=TopRow, fg=black, j}, 
     row{2}  = {bg=NormalRow, fg=black, c}, 
     row{3}  = {bg=NormalRow, fg=black, c}, 
     row{4}  = {bg=NormalRow, fg=black, c}, 
     row{5}  = {bg=NormalRow, fg=black, c}, 
          } 
    \hline
    {\bf Alice's chit} &{\bf Bob's gate}\\
    \hline
     {00}                    &${I}$\\      
       \hline
     {01}                    &${Z}$\\
     \hline
     {10}                    &${X}$\\
     \hline
     {11}                    &${Z X}$\\
     \hline
   \end{tblr}
\caption{Bob's chart. The chart instructs what gate card(s) Bob needs to pick up and submit to QG
depending on what is written on the chit handed to him by Alice.}
\label{tab:bob:chart}
\end{table}
\\
\begin{figure}[!htp]
  \centering
   \begin{subfigure}[b]{0.3\textwidth}
      \centering
      \includegraphics[height=4.5cm]{\FIGDIR/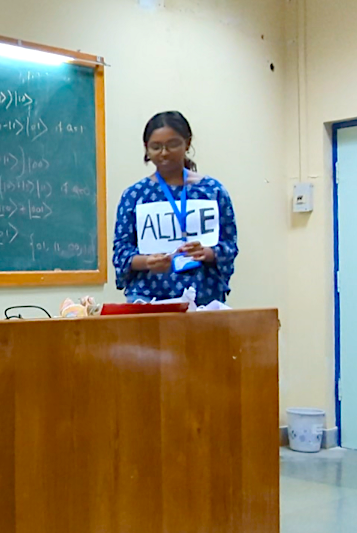}
      \caption{}
      \label{fig:alices:chit:bobs:chart:a}  
    \end{subfigure}
    \begin{subfigure}[b]{0.3\textwidth}
      \centering
      \includegraphics[height=4cm]{\FIGDIR/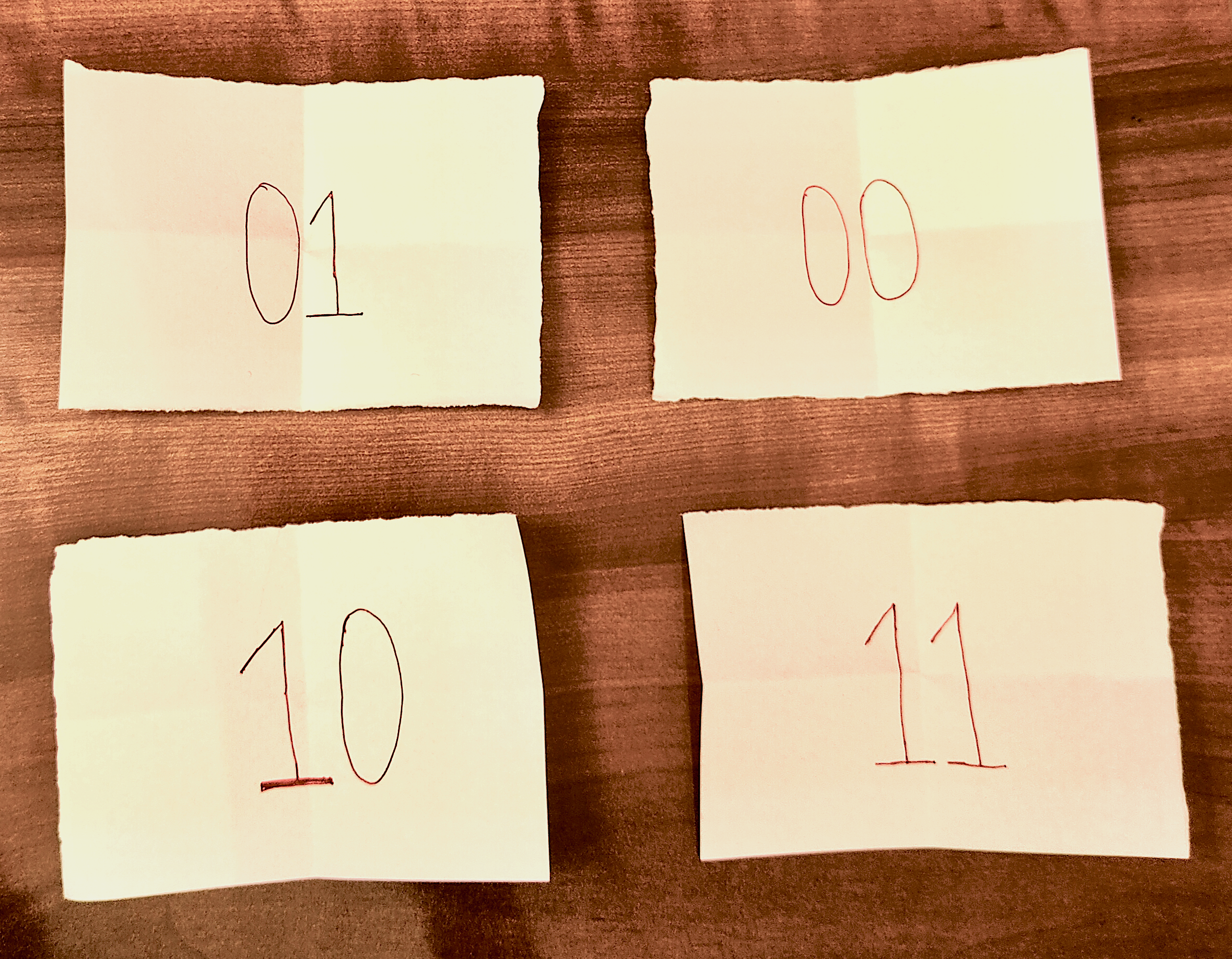}
      \caption{}
      \label{fig:alices:chit:bobs:chart:b}  
    \end{subfigure}
    \begin{subfigure}[b]{0.3\textwidth}
      \centering
       \includegraphics[height=4.5cm]{\FIGDIR/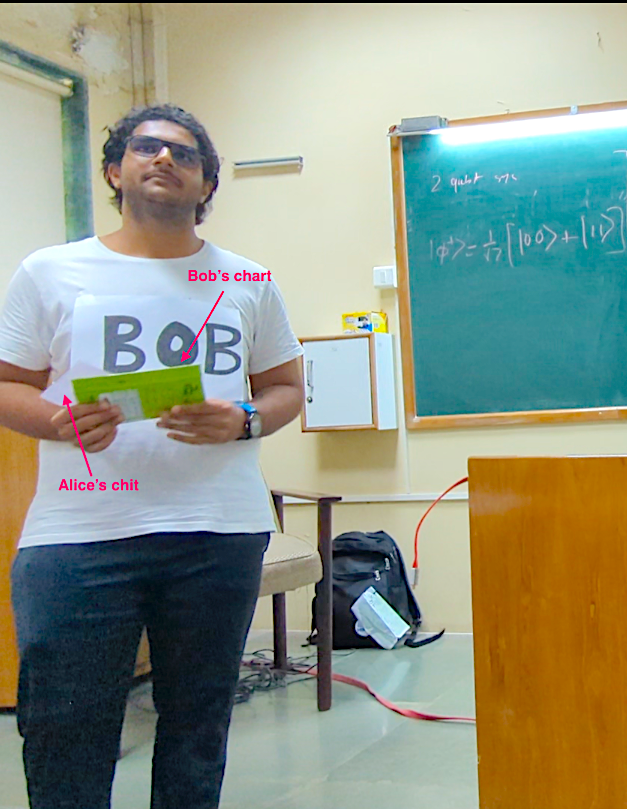}
      \caption{}
    \label{fig:alices:chit:bobs:chart:c}  
    \end{subfigure}
\caption{(a) Alice picks up a chit through a lottery. (b) All of the chits with 4 possible outcomes: `00', `01', `10', `11'.  (c) Bob looks into the information in the chit and applies his gate(s) accordingly.} 
\label{fig:alices:chit:bobs:chart} 
\end{figure}
Following the chart, Bob picks up (i) an $I$-card (identity operation, i.e. no operation), 
(ii) a $Z$-card, (iii) an $X$-card or (iv) first an $X$-card and then a $Z$-card when he finds
$00$, $01$, $10$ or $11$ respectively written on Alice's chit. 
 
%
\i Now,  QG also separately creates a chart according to Alice's measurement outcome (written on the chit). The chart contains the information of Bob's state before he submits any gate cards to QG.
\colorlet{TopRow}{red!80}
\colorlet{NormalRow}{red!20}
\begin{table}[!h]
\centering
    \begin{tblr}{
     colspec = {|c|c|},
     row{1}  = {bg=TopRow, fg=black, j}, 
     row{2}  = {bg=NormalRow, fg=black, c}, 
     row{3}  = {bg=NormalRow, fg=black, c}, 
     row{4}  = {bg=NormalRow, fg=black, c}, 
     row{5}  = {bg=NormalRow, fg=black, c}, 
          } 
    \hline
    {\bf Alice's chit} &{\bf Bob's state}\\
    \hline
     {00}                    & $\ket{1}$\\      
       \hline
     {01}                    & $\ket{0}$\\
     \hline
     {10}                    &  $-\ket{1}$\\
     \hline
     {11}                    & $-\ket{0}$\\
     \hline
   \end{tblr}
\caption{QG's chart for $\ket{Q}=\keto$.}
\label{tab:qg:chart:ketone}
\end{table}
To verify if Bob's gate operation has correctly generated $\ket{Q}$ for his qubit, QG now reveals 
the state to Bob following their chart.   
 \i Bob already knows his gate(s) and now after QG's revelation, he knows his state before his gate operation(s). So he readily checks the final state of his qubit. 
For instance, if Alice's outcome is $11$, Bob's chart reads
$Z X$ and QG's chart reads $-\ket{0}$. Acting $Z X$ on $-\ket{0}$, Bob obtains $\keto$ [$X$-gate first flips his qubit's state and then changes the sign or phase of it]. This state is
the desired state  $\ket{Q}$. QG confirms Bob's final state is indeed $\ket{Q}$, which was originally supplied to Alice by them. Hence, the game ends.   
\end{enumerate}

\ss{What if \boldmath $\ket{Q} = \ket{0}$?}

Now, if QG decides to teleport the other kind of single quantum state, i.e. $\ket{Q} = \ketz$, 
the game rules and steps remain the same. However, QG updates the states
differently in their diary:   
\blgn
{\tcol \ket{\psi_0}}&= \ket{0}\ket{0}\ket{0} = \ket{000}\,.\\
{\tcol \ket{\psi_1}}&= H(1)\ket{\psi_0} 
  =\ket{0}\obst\big[\ketz + \keto\big]\ket{0}\non\\ 
  &= \obst\big[\ket{000} + \ket{010}\big]\non\\
   &\quad\mbox{[after Alice submits $H(1)$ card.]}\\ 
{\tcol \ket{\psi_2}}&= \h X_C(1,2)\ket{\psi_1}
    = \obst\big[\ket{000} + \ket{110}\big]\non\\
       &\quad\mbox{[after Alice submits $CX(1,2)$ card.]}\\ 
{\tcol \ket{\psi_3}}&= \h X_C(0,1)\ket{\psi_2}
                     = \obst\big[\ket{000} + \ket{110}\big]\non\\ 
       &\quad\mbox{[unaltered, after Alice submits $CX(0,1)$ card.]}\\ 
{\tcol \ket{\psi_4}}&= {\h H}(0)\ket{\psi_3}
     =\hf\ket{00}\big[\ket{0}+\ket{1}\big] + \hf\ket{11}\big[(\ket{0}+\ket{1}\big]\non\\ 
      &\quad\mbox{[after Alice submits $H(0)$ card]}\non\\
      &= \hf\big[\ket{000} + \ket{001} + \ket{110} + \ket{111}\big]\,. 
\elgn

Just like in the previous scenario, Alice again picks one classical bit-pair out of the set \{00, 01, 10, 11\} and Bob follows the same chart. However, for the verification of the successful QT, QG provides a different chart to Bob: 
\colorlet{TopRow}{red!80}
\colorlet{NormalRow}{red!20}
\begin{table}[!h]
\centering
    \begin{tblr}{
     colspec = {|c|c|},
     row{1}  = {bg=TopRow, fg=black, j}, 
     row{2}  = {bg=NormalRow, fg=black, c}, 
     row{3}  = {bg=NormalRow, fg=black, c}, 
     row{4}  = {bg=NormalRow, fg=black, c}, 
     row{5}  = {bg=NormalRow, fg=black, c}, 
          } 
    \hline
    {\bf Alice's chit} &{\bf Bob's state}\\
    \hline
     {00}                    & $\ket{0}$\\      
       \hline
     {01}                    & $\ket{0}$\\
     \hline
     {10}                    &  $\ket{1}$\\
     \hline
     {11}                    & $\ket{1}$\\
     \hline
   \end{tblr}
\caption{QG's chart for $\ket{Q}=\ketz$.}
\label{tab:qg:chart:ketzero}
\end{table}
\newline
Note that, in the above, $Z$ gate acts like an identity operator since the operated state remain always in state $\ketz$.

\ss{What if \boldmath $\ket{Q} = a\ketz + b\keto$ ($a\ne 0$, $b\ne 0$) ?}
If QG plans to teleport a generic superposition state
$\ket{Q} = a\ketz + b\keto$, as already discussed in \sref{sec:qt:protocol}, QG notes down
the states that are combinations of the updates for both $\ket{Q} = \ketz$ and $\ket{Q} = \keto$
with appropriate coefficients $a$ and $b$. Instead of mentioning all of them, we mention 
the final update:   

\blgn
{\tcol \ket{\psi_4}} 
            &= \hf\bigg[
            \big[a\ketz + b\keto\big] \ket{00} 
            +\big[a\ketz - b\keto\big] \ket{01}\non\\
            &\quad\big[a\keto + b\ketz\big] \ket{10} 
            +\big[a\keto - b\ketz\big] \ket{11} 
          \bigg]\,.
\elgn
This state is the same as the state described in \eref{eq:qt:protocol:final}.
Following this, QG reveals the following chart to Bob before he applies his gates 
to verify the teleported state.
\colorlet{TopRow}{red!80}
\colorlet{NormalRow}{red!20}
\begin{table}[!h]
\centering
    \begin{tblr}{
     colspec = {|c|c|},
     row{1}  = {bg=TopRow, fg=black, j}, 
     row{2}  = {bg=NormalRow, fg=black, c}, 
     row{3}  = {bg=NormalRow, fg=black, c}, 
     row{4}  = {bg=NormalRow, fg=black, c}, 
     row{5}  = {bg=NormalRow, fg=black, c}, 
          } 
    \hline
    {\bf Alice's chit} &{\bf Bob's state}\\
    \hline
     {00}                    & a$\ket{0} + b\ket{1}$\\      
       \hline
     {01}                    & a$\ket{0} - b\ket{1}$\\
     \hline
     {10}                    & a$\ket{1} + b\ket{0}$\\
     \hline
     {11}                    & a$\ket{1} - b\ket{0}$\\
     \hline
   \end{tblr}
\caption{QG's chart for $\ket{Q}=a\ketz + b\keto$.}
\label{tab:qg:chart:ketgeneric}
\end{table}
\newline
\section{Summary}
In this paper, we discussed a simple game that involves three players to demonstrate quantum teleportation of a single qubit. Quantum teleportation is a complex concept and such demonstration can be useful to engage undergraduate students in learning the basics of it or train teachers who may find it useful in their teaching. Very recently, Nunavath \etal~\cite{nunavath:mishra:pathak:arxiv2408} proposed a \emph{qandies} (quantum candies) model~\cite{lin:mor:proceed20,lin:mor:shapira:arxiv2110} that describes the QT protocol in terms of the candies' classical entities (e.g. colors and tastes). In principle, instead of the `quantum' chits can be replaced by qandies, that Alice picks up during her measurement through a lottery. Our kind of game can be  modified and played for other entanglement based protocols such as superdense coding~\cite{bennet:wiesner:prl92}, entanglement swapping~\cite{zukowski:zeilinger:horne:ekert:prl93}, and quantum key distribution~\cite{bennett:brassard:proceed84,ekert:prl91}.

\begin{acknowledgments}
HB thanks Ananya Vinod (played the role of Alice), Daksh Gupta (played Bob), Arkaprava Bose (played Quantum God), and other NIUS students without whose enthusiastic participation, the game could not have been conducted successfully and this article would not have seen its final version. He also thanks Karthik Shetty and Mamatha Maddur from HBCSE, Mumbai, for providing the necessary equipment. Ananya, Karthik, and Spandan Mandal helped him by providing with some needful images. The author is also indebted to Dr Praveen Pathak, the organizer of the NIUS Physics 2024 camp at HBCSE and Dr Deepak Garg for taking part in very engaging discussions during the planning and demonstration of the game. Finally, he thanks IBM for its open-source \emph{Qiskit} SDK, which is used in generating quantum circuit diagrams discussed in the paper.    
\end{acknowledgments}
\appendix
\section{Basic quantum gates}
\label{app:basic:qgates}
We briefly discuss the gates that have been used in the QT protocol discussed in the main part of the paper. 
All of these gates can be represented by $2\times 2$ unitary Hermitian matrices which act on the single qubit bases
$\ketz = \bbmat 1\\ 0 \ebmat$ and $\keto = \bbmat 0\\ 1 \ebmat$ during the protocol operations. Note that in this basis set, a generic $2\times 2$ matrix $\h M$ with elements $\alpha$, $\beta$, $\gamma$, and $\delta$ can 
be expressed as
\blgn
M= 
\bbmat 
\alpha & \beta \\ 
\gamma & \delta 
\ebmat
= \al\ketz\braz + \bt\ketz\brao +\g\keto\braz +\del\keto\brao\,.
\elgn
\def\hal{{\h \alpha}}
\def\hbt{{\h \beta}}
\def\hgm{{\h \gamma}}
\def\hdl{{\h \delta}}
If the matrix elements become individual matrices themselves (e.g. $\hal, \hbt, \hgm, \hdl$), we have
\blgn
\tilde{M} &= 
\bbmat 
\hal & \hbt \\ 
\hgm & \hdl 
\ebmat
= \hal\otimes\ketz\braz + \hbt\otimes\ketz\brao \non\\
&\quad+\hgm\otimes\keto\braz +\hdl\otimes\keto\brao\,.
\elgn
We follow the above two identities to express gate matrices in the discussion below. 

\subsection{Pauli gates}
The four Pauli matrices act as gates on a single qubit and they are represented as
\blgn
\h X &= \bbmat 0 & 1\\ 1 & 0 \ebmat = \ketz\brao + \keto\braz\,.\label{eq:x}\\
\h Y &= \bbmat 0 & -i\\ i & 0 \ebmat = i\big[\ketz\brao - \keto\braz]\,.\\
\h Z &= \bbmat 1 & 0\\ 0 & -1 \ebmat = \ketz\braz - \keto\brao\,.\label{eq:z}\\
\h I &= \bbmat 1 & 0\\ 0 & 1 \ebmat = \ketz\braz + \keto\brao\,.
\elgn
The above operators become $X$, $Y$, $Z$, and $I$ gates. It is easy to see that an $X$ gate acts as a bit-flip or \emph{NOT} gate as it changes qubit $\ketz$ to $\keto$ and vice-versa. $Y$ gate (not part of the QT protocol), also flips the bit, however, it picks up a  negative sign ($e^{i\pi}$ phase) when operated on $\keto$. Similarly, the identity Pauli matrix acts as an $I$ gate, causing no changes in the quantum states while $Z$ gate changes the sign of the amplitude of $\keto$ when acted on it.

\subsection{Hadamard ($H$) gate}
The Hadamard or $H$ gate is a combination of $X$ and $Z$ gates with a normalization factor ($1/\sqrt{2})$). In operator form:
\blgn
\h H = \obst\big[\h X + \h Z\big] =\obst \bbmat 1 & 1\\ 1 & -1 \ebmat\,. 
\elgn 
From \eref{eq:x} and \eref{eq:z}, one can write the operator in terms of outer product form
\blgn
\h H = \obst\bigg[ {\tcoll\big[\ketz + \keto\big]}\braz + {\tcoll\big[\ketz - \keto\big]}\brao\bigg]\,.  
\elgn
Thus, $H$ gate creates a superposition state $\ketz \pm \keto$, with the normalization factor $1/\sqrt{2}$, when acted on $\ketz$ and $\keto$, respectively.

\subsection{Controlled gates}
$CX$ and $CZ$ gates stand for \emph{controlled} $X$ and $Z$ gates respectively and they are two-qubit gates.  
 In a two-qubit state, if the first qubit (\emph{control bit}) is in state $\keto$, then a controlled gate operates on the second qubit (\emph{target bit}). The indices of the qubits are mentioned orderwise in the parentheses after the name of the gate. Operators representing $CX(0,1)$ and $CZ(0,1)$ gates are defined as (here $0$ is the first/control bit, $1$ is the second/target bit)  
%
\begin{figure}[!htp]
  \centering
   \begin{subfigure}[b]{0.3\textwidth}
      \centering
      \includegraphics[height=1.5cm,clip]{\FIGDIR/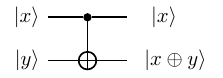}
      \caption{}
      \label{fig:cx:gate}
    \end{subfigure}
    \begin{subfigure}[b]{0.3\textwidth}
      \centering
      \includegraphics[height=1.5cm,clip]{\FIGDIR/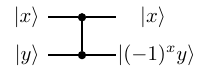}
      \caption{}
      \label{fig:cz:gate}
    \end{subfigure}   
\caption{(a) $CX$ and (b) $CZ$ gates for input qubits $\ket{x}$ and $\ket{y}$. One of the outputs remains the same as one of the inputs (control bit $\ket{x}$) in both gates. After  $CX$ gate operation the other output for the target bit $\ket{y}$ becomes the output of a classical $XOR$ gate. In case of $CZ$ operation, the target bit's output conditionally changes its sign: $\ket{y} \to -\ket{y}$ only when $\ket{x}=\keto$.} 
\label{fig:cx:cz:gates}
\end{figure} 
\blgn
\h X_C(0,1) &= \bbmat {\h I} & 0 \\ 0 & {\h X} \ebmat
    \non\\
    &\quad = {\h I}\otimes\ketz\braz + {\h X}\otimes\keto\brao    \non\\
    &\quad = \ketzz\brazz + \ketoz\braoz + \ketzo\braoo + \ketoo\brazo\,.\\
\h X_Z(0,1) &= \bbmat {\h I} & 0 \\ 0 & {\h Z} \ebmat
    \non\\
    &\quad = {\h I}\otimes\ketz\braz + {\h Z}\otimes\keto\brao \non\\
    &\quad = \ketzz\brazz + \ketoz\braoz + \ketzo\brazo - \ketoo\braoo\,.
\elgn
Note that the first bit after the $CX$ operation remains unchanged while 
the second output becomes the classical $XOR$ output of the two input bits. This justifies the $XOR$ symbol in the control bit node of the diagrammatic representation (see \fref{fig:cx:gate}). On the other hand, $CZ$ allows sign change for the second bit only when the first bit is $1$, lacking a classical gate analog. One can show $CZ(0,1) = CZ(1,0)$, i.e. the control and target bits are interchangeable and hence both bit-nodes are represented by the control-node symbol (filled circle, see \fref{fig:cz:gate}).      
   
\bibliographystyle{apsrev4-1}
\bibliography{refs_qt_game_by_hbar.bib}

\begin{thebibliography}{40}%
\makeatletter
\providecommand \@ifxundefined [1]{%
 \@ifx{#1\undefined}
}%
\providecommand \@ifnum [1]{%
 \ifnum #1\expandafter \@firstoftwo
 \else \expandafter \@secondoftwo
 \fi
}%
\providecommand \@ifx [1]{%
 \ifx #1\expandafter \@firstoftwo
 \else \expandafter \@secondoftwo
 \fi
}%
\providecommand \natexlab [1]{#1}%
\providecommand \enquote  [1]{``#1''}%
\providecommand \bibnamefont  [1]{#1}%
\providecommand \bibfnamefont [1]{#1}%
\providecommand \citenamefont [1]{#1}%
\providecommand \href@noop [0]{\@secondoftwo}%
\providecommand \href [0]{\begingroup \@sanitize@url \@href}%
\providecommand \@href[1]{\@@startlink{#1}\@@href}%
\providecommand \@@href[1]{\endgroup#1\@@endlink}%
\providecommand \@sanitize@url [0]{\catcode `\\12\catcode `\$12\catcode
  `\&12\catcode `\#12\catcode `\^12\catcode `\_12\catcode `\%12\relax}%
\providecommand \@@startlink[1]{}%
\providecommand \@@endlink[0]{}%
\providecommand \url  [0]{\begingroup\@sanitize@url \@url }%
\providecommand \@url [1]{\endgroup\@href {#1}{\urlprefix }}%
\providecommand \urlprefix  [0]{URL }%
\providecommand \Eprint [0]{\href }%
\providecommand \doibase [0]{http://dx.doi.org/}%
\providecommand \selectlanguage [0]{\@gobble}%
\providecommand \bibinfo  [0]{\@secondoftwo}%
\providecommand \bibfield  [0]{\@secondoftwo}%
\providecommand \translation [1]{[#1]}%
\providecommand \BibitemOpen [0]{}%
\providecommand \bibitemStop [0]{}%
\providecommand \bibitemNoStop [0]{.\EOS\space}%
\providecommand \EOS [0]{\spacefactor3000\relax}%
\providecommand \BibitemShut  [1]{\csname bibitem#1\endcsname}%
\let\auto@bib@innerbib\@empty
\bibitem [{ray()}]{ray:ggbb:69}%
  \BibitemOpen
  \href@noop {} {}\bibinfo {howpublished}
  {\url{https://www.imdb.com/title/tt0063023/}}\BibitemShut {NoStop}%
\bibitem [{\citenamefont {Bennett}\ \emph {et~al.}(1993)\citenamefont
  {Bennett}, \citenamefont {Brassard}, \citenamefont {Cr\'epeau}, \citenamefont
  {Jozsa}, \citenamefont {Peres},\ and\ \citenamefont
  {Wootters}}]{bennett:etal:prl93}%
  \BibitemOpen
  \bibfield  {author} {\bibinfo {author} {\bibfnamefont {C.~H.}\ \bibnamefont
  {Bennett}}, \bibinfo {author} {\bibfnamefont {G.}~\bibnamefont {Brassard}},
  \bibinfo {author} {\bibfnamefont {C.}~\bibnamefont {Cr\'epeau}}, \bibinfo
  {author} {\bibfnamefont {R.}~\bibnamefont {Jozsa}}, \bibinfo {author}
  {\bibfnamefont {A.}~\bibnamefont {Peres}}, \ and\ \bibinfo {author}
  {\bibfnamefont {W.~K.}\ \bibnamefont {Wootters}},\ }\href
  {https://link.aps.org/doi/10.1103/PhysRevLett.70.1895} {\bibfield  {journal}
  {\bibinfo  {journal} {Phys. Rev. Lett.}\ }\textbf {\bibinfo {volume} {70}},\
  \bibinfo {pages} {1895} (\bibinfo {year} {1993})}\BibitemShut {NoStop}%
\bibitem [{\citenamefont {Bouwmeester}\ \emph {et~al.}(1997)\citenamefont
  {Bouwmeester}, \citenamefont {Pan}, \citenamefont {Mattle}, \citenamefont
  {Eibl}, \citenamefont {Weinfurter},\ and\ \citenamefont
  {Zeilinger}}]{bouwmeester:etal:zeilinger:gr:nat97}%
  \BibitemOpen
  \bibfield  {author} {\bibinfo {author} {\bibfnamefont {D.}~\bibnamefont
  {Bouwmeester}}, \bibinfo {author} {\bibfnamefont {J.-W.}\ \bibnamefont
  {Pan}}, \bibinfo {author} {\bibfnamefont {K.}~\bibnamefont {Mattle}},
  \bibinfo {author} {\bibfnamefont {M.}~\bibnamefont {Eibl}}, \bibinfo {author}
  {\bibfnamefont {H.}~\bibnamefont {Weinfurter}}, \ and\ \bibinfo {author}
  {\bibfnamefont {A.}~\bibnamefont {Zeilinger}},\ }\href {\doibase
  10.1038/37539} {\bibfield  {journal} {\bibinfo  {journal} {Nature}\ }\textbf
  {\bibinfo {volume} {390}},\ \bibinfo {pages} {575} (\bibinfo {year}
  {1997})}\BibitemShut {NoStop}%
\bibitem [{\citenamefont {Braunstein}\ and\ \citenamefont
  {Kimble}(1998)}]{braunstein:kimble:prl98}%
  \BibitemOpen
  \bibfield  {author} {\bibinfo {author} {\bibfnamefont {S.~L.}\ \bibnamefont
  {Braunstein}}\ and\ \bibinfo {author} {\bibfnamefont {H.~J.}\ \bibnamefont
  {Kimble}},\ }\href {\doibase 10.1103/PhysRevLett.80.869} {\bibfield
  {journal} {\bibinfo  {journal} {Phys. Rev. Lett.}\ }\textbf {\bibinfo
  {volume} {80}},\ \bibinfo {pages} {869} (\bibinfo {year} {1998})}\BibitemShut
  {NoStop}%
\bibitem [{\citenamefont {Nielsen}\ \emph {et~al.}(1998)\citenamefont
  {Nielsen}, \citenamefont {Knill},\ and\ \citenamefont
  {Laflamme}}]{nielsen:laflamme:nat98}%
  \BibitemOpen
  \bibfield  {author} {\bibinfo {author} {\bibfnamefont {M.~A.}\ \bibnamefont
  {Nielsen}}, \bibinfo {author} {\bibfnamefont {E.}~\bibnamefont {Knill}}, \
  and\ \bibinfo {author} {\bibfnamefont {R.}~\bibnamefont {Laflamme}},\ }\href
  {\doibase 10.1038/23891} {\bibfield  {journal} {\bibinfo  {journal} {Nature}\
  }\textbf {\bibinfo {volume} {396}},\ \bibinfo {pages} {52} (\bibinfo {year}
  {1998})}\BibitemShut {NoStop}%
\bibitem [{\citenamefont {Furusawa}\ \emph {et~al.}(1998)\citenamefont
  {Furusawa}, \citenamefont {Sørensen}, \citenamefont {Braunstein},
  \citenamefont {Fuchs}, \citenamefont {Kimble},\ and\ \citenamefont
  {Polzik}}]{furusawa:etal:sc98}%
  \BibitemOpen
  \bibfield  {author} {\bibinfo {author} {\bibfnamefont {A.}~\bibnamefont
  {Furusawa}}, \bibinfo {author} {\bibfnamefont {J.~L.}\ \bibnamefont
  {Sørensen}}, \bibinfo {author} {\bibfnamefont {S.~L.}\ \bibnamefont
  {Braunstein}}, \bibinfo {author} {\bibfnamefont {C.~A.}\ \bibnamefont
  {Fuchs}}, \bibinfo {author} {\bibfnamefont {H.~J.}\ \bibnamefont {Kimble}}, \
  and\ \bibinfo {author} {\bibfnamefont {E.~S.}\ \bibnamefont {Polzik}},\
  }\href {\doibase 10.1126/science.282.5389.706} {\bibfield  {journal}
  {\bibinfo  {journal} {Science}\ }\textbf {\bibinfo {volume} {282}},\ \bibinfo
  {pages} {706} (\bibinfo {year} {1998})}\BibitemShut {NoStop}%
\bibitem [{\citenamefont {Takei}\ \emph {et~al.}(2005)\citenamefont {Takei},
  \citenamefont {Aoki}, \citenamefont {Koike}, \citenamefont {Yoshino},
  \citenamefont {Wakui}, \citenamefont {Yonezawa}, \citenamefont {Hiraoka},
  \citenamefont {Mizuno}, \citenamefont {Takeoka}, \citenamefont {Ban},\ and\
  \citenamefont {Furusawa}}]{takei:etal:pra05}%
  \BibitemOpen
  \bibfield  {author} {\bibinfo {author} {\bibfnamefont {N.}~\bibnamefont
  {Takei}}, \bibinfo {author} {\bibfnamefont {T.}~\bibnamefont {Aoki}},
  \bibinfo {author} {\bibfnamefont {S.}~\bibnamefont {Koike}}, \bibinfo
  {author} {\bibfnamefont {K.-i.}\ \bibnamefont {Yoshino}}, \bibinfo {author}
  {\bibfnamefont {K.}~\bibnamefont {Wakui}}, \bibinfo {author} {\bibfnamefont
  {H.}~\bibnamefont {Yonezawa}}, \bibinfo {author} {\bibfnamefont
  {T.}~\bibnamefont {Hiraoka}}, \bibinfo {author} {\bibfnamefont
  {J.}~\bibnamefont {Mizuno}}, \bibinfo {author} {\bibfnamefont
  {M.}~\bibnamefont {Takeoka}}, \bibinfo {author} {\bibfnamefont
  {M.}~\bibnamefont {Ban}}, \ and\ \bibinfo {author} {\bibfnamefont
  {A.}~\bibnamefont {Furusawa}},\ }\href {\doibase 10.1103/PhysRevA.72.042304}
  {\bibfield  {journal} {\bibinfo  {journal} {Phys. Rev. A}\ }\textbf {\bibinfo
  {volume} {72}},\ \bibinfo {pages} {042304} (\bibinfo {year}
  {2005})}\BibitemShut {NoStop}%
\bibitem [{\citenamefont {Riebe}\ \emph {et~al.}(2004)\citenamefont {Riebe},
  \citenamefont {H{\"a}ffner}, \citenamefont {Roos}, \citenamefont
  {H{\"a}nsel}, \citenamefont {Benhelm}, \citenamefont {Lancaster},
  \citenamefont {K{\"o}rber}, \citenamefont {Becher}, \citenamefont
  {Schmidt-Kaler}, \citenamefont {James},\ and\ \citenamefont
  {Blatt}}]{riebe:etal:nat04}%
  \BibitemOpen
  \bibfield  {author} {\bibinfo {author} {\bibfnamefont {M.}~\bibnamefont
  {Riebe}}, \bibinfo {author} {\bibfnamefont {H.}~\bibnamefont {H{\"a}ffner}},
  \bibinfo {author} {\bibfnamefont {C.~F.}\ \bibnamefont {Roos}}, \bibinfo
  {author} {\bibfnamefont {W.}~\bibnamefont {H{\"a}nsel}}, \bibinfo {author}
  {\bibfnamefont {J.}~\bibnamefont {Benhelm}}, \bibinfo {author} {\bibfnamefont
  {G.~P.~T.}\ \bibnamefont {Lancaster}}, \bibinfo {author} {\bibfnamefont
  {T.~W.}\ \bibnamefont {K{\"o}rber}}, \bibinfo {author} {\bibfnamefont
  {C.}~\bibnamefont {Becher}}, \bibinfo {author} {\bibfnamefont
  {F.}~\bibnamefont {Schmidt-Kaler}}, \bibinfo {author} {\bibfnamefont
  {D.~F.~V.}\ \bibnamefont {James}}, \ and\ \bibinfo {author} {\bibfnamefont
  {R.}~\bibnamefont {Blatt}},\ }\href {\doibase 10.1038/nature02570} {\bibfield
   {journal} {\bibinfo  {journal} {Nature}\ }\textbf {\bibinfo {volume}
  {429}},\ \bibinfo {pages} {734} (\bibinfo {year} {2004})}\BibitemShut
  {NoStop}%
\bibitem [{\citenamefont {Barrett}\ \emph {et~al.}(2004)\citenamefont
  {Barrett}, \citenamefont {Chiaverini}, \citenamefont {Schaetz}, \citenamefont
  {Britton}, \citenamefont {Itano}, \citenamefont {Jost}, \citenamefont
  {Knill}, \citenamefont {Langer}, \citenamefont {Leibfried}, \citenamefont
  {Ozeri},\ and\ \citenamefont {Wineland}}]{barrett:etal:nat04}%
  \BibitemOpen
  \bibfield  {author} {\bibinfo {author} {\bibfnamefont {M.~D.}\ \bibnamefont
  {Barrett}}, \bibinfo {author} {\bibfnamefont {J.}~\bibnamefont {Chiaverini}},
  \bibinfo {author} {\bibfnamefont {T.}~\bibnamefont {Schaetz}}, \bibinfo
  {author} {\bibfnamefont {J.}~\bibnamefont {Britton}}, \bibinfo {author}
  {\bibfnamefont {W.~M.}\ \bibnamefont {Itano}}, \bibinfo {author}
  {\bibfnamefont {J.~D.}\ \bibnamefont {Jost}}, \bibinfo {author}
  {\bibfnamefont {E.}~\bibnamefont {Knill}}, \bibinfo {author} {\bibfnamefont
  {C.}~\bibnamefont {Langer}}, \bibinfo {author} {\bibfnamefont
  {D.}~\bibnamefont {Leibfried}}, \bibinfo {author} {\bibfnamefont
  {R.}~\bibnamefont {Ozeri}}, \ and\ \bibinfo {author} {\bibfnamefont {D.~J.}\
  \bibnamefont {Wineland}},\ }\href {\doibase 10.1038/nature02608} {\bibfield
  {journal} {\bibinfo  {journal} {Nature}\ }\textbf {\bibinfo {volume} {429}},\
  \bibinfo {pages} {737} (\bibinfo {year} {2004})}\BibitemShut {NoStop}%
\bibitem [{\citenamefont {Wan}\ \emph {et~al.}(2019)\citenamefont {Wan} \emph
  {et~al.}}]{wan:etal:sc19}%
  \BibitemOpen
  \bibfield  {author} {\bibinfo {author} {\bibfnamefont {Y.}~\bibnamefont
  {Wan}} \emph {et~al.},\ }\href {\doibase 10.1126/science.aaw9415} {\bibfield
  {journal} {\bibinfo  {journal} {Science}\ }\textbf {\bibinfo {volume}
  {364}},\ \bibinfo {pages} {875} (\bibinfo {year} {2019})}\BibitemShut
  {NoStop}%
\bibitem [{\citenamefont {Sherson}\ \emph {et~al.}(2006)\citenamefont
  {Sherson}, \citenamefont {Krauter}, \citenamefont {Olsson}, \citenamefont
  {Julsgaard}, \citenamefont {Hammerer}, \citenamefont {Cirac},\ and\
  \citenamefont {Polzik}}]{sherson:etal:nat06}%
  \BibitemOpen
  \bibfield  {author} {\bibinfo {author} {\bibfnamefont {J.~F.}\ \bibnamefont
  {Sherson}}, \bibinfo {author} {\bibfnamefont {H.}~\bibnamefont {Krauter}},
  \bibinfo {author} {\bibfnamefont {R.~K.}\ \bibnamefont {Olsson}}, \bibinfo
  {author} {\bibfnamefont {B.}~\bibnamefont {Julsgaard}}, \bibinfo {author}
  {\bibfnamefont {K.}~\bibnamefont {Hammerer}}, \bibinfo {author}
  {\bibfnamefont {I.}~\bibnamefont {Cirac}}, \ and\ \bibinfo {author}
  {\bibfnamefont {E.~S.}\ \bibnamefont {Polzik}},\ }\href {\doibase
  10.1038/nature05136} {\bibfield  {journal} {\bibinfo  {journal} {Nature}\
  }\textbf {\bibinfo {volume} {443}},\ \bibinfo {pages} {557} (\bibinfo {year}
  {2006})}\BibitemShut {NoStop}%
\bibitem [{\citenamefont {Pfaff}\ \emph {et~al.}(2014)\citenamefont {Pfaff}
  \emph {et~al.}}]{pfaff:etal:sc14}%
  \BibitemOpen
  \bibfield  {author} {\bibinfo {author} {\bibnamefont {Pfaff}} \emph
  {et~al.},\ }\href {\doibase 10.1126/science.125351} {\bibfield  {journal}
  {\bibinfo  {journal} {Science}\ }\textbf {\bibinfo {volume} {345}},\ \bibinfo
  {pages} {eaau1255} (\bibinfo {year} {2014})}\BibitemShut {NoStop}%
\bibitem [{\citenamefont {Reindl}\ \emph {et~al.}(2018)\citenamefont {Reindl}
  \emph {et~al.}}]{reindl:etal:sciadv18}%
  \BibitemOpen
  \bibfield  {author} {\bibinfo {author} {\bibfnamefont {M.}~\bibnamefont
  {Reindl}} \emph {et~al.},\ }\href {\doibase 10.1126/sciadv.aau1255}
  {\bibfield  {journal} {\bibinfo  {journal} {Science Advances}\ }\textbf
  {\bibinfo {volume} {4}},\ \bibinfo {pages} {eaau1255} (\bibinfo {year}
  {2018})}\BibitemShut {NoStop}%
\bibitem [{\citenamefont {Steffen}\ \emph {et~al.}(2013)\citenamefont
  {Steffen}, \citenamefont {Salathe}, \citenamefont {Oppliger}, \citenamefont
  {Kurpiers}, \citenamefont {Baur}, \citenamefont {Lang}, \citenamefont
  {Eichler}, \citenamefont {Puebla-Hellmann}, \citenamefont {Fedorov},\ and\
  \citenamefont {Wallraff}}]{steffen:etal:nat13}%
  \BibitemOpen
  \bibfield  {author} {\bibinfo {author} {\bibfnamefont {L.}~\bibnamefont
  {Steffen}}, \bibinfo {author} {\bibfnamefont {Y.}~\bibnamefont {Salathe}},
  \bibinfo {author} {\bibfnamefont {M.}~\bibnamefont {Oppliger}}, \bibinfo
  {author} {\bibfnamefont {P.}~\bibnamefont {Kurpiers}}, \bibinfo {author}
  {\bibfnamefont {M.}~\bibnamefont {Baur}}, \bibinfo {author} {\bibfnamefont
  {C.}~\bibnamefont {Lang}}, \bibinfo {author} {\bibfnamefont {C.}~\bibnamefont
  {Eichler}}, \bibinfo {author} {\bibfnamefont {G.}~\bibnamefont
  {Puebla-Hellmann}}, \bibinfo {author} {\bibfnamefont {A.}~\bibnamefont
  {Fedorov}}, \ and\ \bibinfo {author} {\bibfnamefont {A.}~\bibnamefont
  {Wallraff}},\ }\href {\doibase 10.1038/nature12422} {\bibfield  {journal}
  {\bibinfo  {journal} {Nature}\ }\textbf {\bibinfo {volume} {500}},\ \bibinfo
  {pages} {319} (\bibinfo {year} {2013})}\BibitemShut {NoStop}%
\bibitem [{\citenamefont {Pirandola}\ \emph {et~al.}(2015)\citenamefont
  {Pirandola}, \citenamefont {Eisert}, \citenamefont {Weedbrook}, \citenamefont
  {Furusawa},\ and\ \citenamefont {Braunstein}}]{pirandola:nphoton15}%
  \BibitemOpen
  \bibfield  {author} {\bibinfo {author} {\bibfnamefont {S.}~\bibnamefont
  {Pirandola}}, \bibinfo {author} {\bibfnamefont {J.}~\bibnamefont {Eisert}},
  \bibinfo {author} {\bibfnamefont {C.}~\bibnamefont {Weedbrook}}, \bibinfo
  {author} {\bibfnamefont {A.}~\bibnamefont {Furusawa}}, \ and\ \bibinfo
  {author} {\bibfnamefont {S.~L.}\ \bibnamefont {Braunstein}},\ }\href
  {\doibase 10.1038/nphoton.2015.154} {\bibfield  {journal} {\bibinfo
  {journal} {Nature Photonics}\ }\textbf {\bibinfo {volume} {9}},\ \bibinfo
  {pages} {641} (\bibinfo {year} {2015})}\BibitemShut {NoStop}%
\bibitem [{\citenamefont {Hu}\ \emph {et~al.}(2023)\citenamefont {Hu},
  \citenamefont {Guo}, \citenamefont {Liu}, \citenamefont {Li},\ and\
  \citenamefont {Guo}}]{hu:etal:nrev23}%
  \BibitemOpen
  \bibfield  {author} {\bibinfo {author} {\bibfnamefont {X.-M.}\ \bibnamefont
  {Hu}}, \bibinfo {author} {\bibfnamefont {Y.}~\bibnamefont {Guo}}, \bibinfo
  {author} {\bibfnamefont {B.-H.}\ \bibnamefont {Liu}}, \bibinfo {author}
  {\bibfnamefont {C.-F.}\ \bibnamefont {Li}}, \ and\ \bibinfo {author}
  {\bibfnamefont {G.-C.}\ \bibnamefont {Guo}},\ }\href {\doibase
  10.1038/s42254-023-00588-x} {\bibfield  {journal} {\bibinfo  {journal}
  {Nature Reviews Physics}\ }\textbf {\bibinfo {volume} {5}},\ \bibinfo {pages}
  {339} (\bibinfo {year} {2023})}\BibitemShut {NoStop}%
\bibitem [{\citenamefont {Bouwmeester}\ \emph {et~al.}(1999)\citenamefont
  {Bouwmeester}, \citenamefont {Pan}, \citenamefont {Daniell}, \citenamefont
  {Weinfurter},\ and\ \citenamefont
  {Zeilinger}}]{boumeester:etal:zeilinger:gr:prl99}%
  \BibitemOpen
  \bibfield  {author} {\bibinfo {author} {\bibfnamefont {D.}~\bibnamefont
  {Bouwmeester}}, \bibinfo {author} {\bibfnamefont {J.-W.}\ \bibnamefont
  {Pan}}, \bibinfo {author} {\bibfnamefont {M.}~\bibnamefont {Daniell}},
  \bibinfo {author} {\bibfnamefont {H.}~\bibnamefont {Weinfurter}}, \ and\
  \bibinfo {author} {\bibfnamefont {A.}~\bibnamefont {Zeilinger}},\ }\href
  {\doibase 10.1103/PhysRevLett.82.1345} {\bibfield  {journal} {\bibinfo
  {journal} {Phys. Rev. Lett.}\ }\textbf {\bibinfo {volume} {82}},\ \bibinfo
  {pages} {1345} (\bibinfo {year} {1999})}\BibitemShut {NoStop}%
\bibitem [{\citenamefont {Pan}\ \emph {et~al.}(2001)\citenamefont {Pan},
  \citenamefont {Daniell}, \citenamefont {Gasparoni}, \citenamefont {Weihs},\
  and\ \citenamefont {Zeilinger}}]{pan:etal:zeilinger:gr:prl01}%
  \BibitemOpen
  \bibfield  {author} {\bibinfo {author} {\bibfnamefont {J.-W.}\ \bibnamefont
  {Pan}}, \bibinfo {author} {\bibfnamefont {M.}~\bibnamefont {Daniell}},
  \bibinfo {author} {\bibfnamefont {S.}~\bibnamefont {Gasparoni}}, \bibinfo
  {author} {\bibfnamefont {G.}~\bibnamefont {Weihs}}, \ and\ \bibinfo {author}
  {\bibfnamefont {A.}~\bibnamefont {Zeilinger}},\ }\href {\doibase
  10.1103/PhysRevLett.86.4435} {\bibfield  {journal} {\bibinfo  {journal}
  {Phys. Rev. Lett.}\ }\textbf {\bibinfo {volume} {86}},\ \bibinfo {pages}
  {4435} (\bibinfo {year} {2001})}\BibitemShut {NoStop}%
\bibitem [{\citenamefont {Zhao}\ \emph {et~al.}(2004)\citenamefont {Zhao},
  \citenamefont {Chen}, \citenamefont {Zhang}, \citenamefont {Yang},
  \citenamefont {Briegel},\ and\ \citenamefont {Pan}}]{zhao:etal:nat04}%
  \BibitemOpen
  \bibfield  {author} {\bibinfo {author} {\bibfnamefont {Z.}~\bibnamefont
  {Zhao}}, \bibinfo {author} {\bibfnamefont {Y.-A.}\ \bibnamefont {Chen}},
  \bibinfo {author} {\bibfnamefont {A.-N.}\ \bibnamefont {Zhang}}, \bibinfo
  {author} {\bibfnamefont {T.}~\bibnamefont {Yang}}, \bibinfo {author}
  {\bibfnamefont {H.~J.}\ \bibnamefont {Briegel}}, \ and\ \bibinfo {author}
  {\bibfnamefont {J.-W.}\ \bibnamefont {Pan}},\ }\href {\doibase
  10.1038/nature02643} {\bibfield  {journal} {\bibinfo  {journal} {Nature}\
  }\textbf {\bibinfo {volume} {430}},\ \bibinfo {pages} {54} (\bibinfo {year}
  {2004})}\BibitemShut {NoStop}%
\bibitem [{\citenamefont {Mafu}\ \emph {et~al.}(2013)\citenamefont {Mafu},
  \citenamefont {Dudley}, \citenamefont {Goyal}, \citenamefont {Giovannini},
  \citenamefont {McLaren}, \citenamefont {Padgett}, \citenamefont {Konrad},
  \citenamefont {Petruccione}, \citenamefont {L\"utkenhaus},\ and\
  \citenamefont {Forbes}}]{mafu:etal:pra13}%
  \BibitemOpen
  \bibfield  {author} {\bibinfo {author} {\bibfnamefont {M.}~\bibnamefont
  {Mafu}}, \bibinfo {author} {\bibfnamefont {A.}~\bibnamefont {Dudley}},
  \bibinfo {author} {\bibfnamefont {S.}~\bibnamefont {Goyal}}, \bibinfo
  {author} {\bibfnamefont {D.}~\bibnamefont {Giovannini}}, \bibinfo {author}
  {\bibfnamefont {M.}~\bibnamefont {McLaren}}, \bibinfo {author} {\bibfnamefont
  {M.~J.}\ \bibnamefont {Padgett}}, \bibinfo {author} {\bibfnamefont
  {T.}~\bibnamefont {Konrad}}, \bibinfo {author} {\bibfnamefont
  {F.}~\bibnamefont {Petruccione}}, \bibinfo {author} {\bibfnamefont
  {N.}~\bibnamefont {L\"utkenhaus}}, \ and\ \bibinfo {author} {\bibfnamefont
  {A.}~\bibnamefont {Forbes}},\ }\href {\doibase 10.1103/PhysRevA.88.032305}
  {\bibfield  {journal} {\bibinfo  {journal} {Phys. Rev. A}\ }\textbf {\bibinfo
  {volume} {88}},\ \bibinfo {pages} {032305} (\bibinfo {year}
  {2013})}\BibitemShut {NoStop}%
\bibitem [{\citenamefont {Luo}\ \emph {et~al.}(2019)\citenamefont {Luo},
  \citenamefont {Zhong}, \citenamefont {Erhard}, \citenamefont {Wang},
  \citenamefont {Peng}, \citenamefont {Krenn}, \citenamefont {Jiang},
  \citenamefont {Li}, \citenamefont {Liu}, \citenamefont {Lu}, \citenamefont
  {Zeilinger},\ and\ \citenamefont {Pan}}]{luo:etal:prl19}%
  \BibitemOpen
  \bibfield  {author} {\bibinfo {author} {\bibfnamefont {Y.-H.}\ \bibnamefont
  {Luo}}, \bibinfo {author} {\bibfnamefont {H.-S.}\ \bibnamefont {Zhong}},
  \bibinfo {author} {\bibfnamefont {M.}~\bibnamefont {Erhard}}, \bibinfo
  {author} {\bibfnamefont {X.-L.}\ \bibnamefont {Wang}}, \bibinfo {author}
  {\bibfnamefont {L.-C.}\ \bibnamefont {Peng}}, \bibinfo {author}
  {\bibfnamefont {M.}~\bibnamefont {Krenn}}, \bibinfo {author} {\bibfnamefont
  {X.}~\bibnamefont {Jiang}}, \bibinfo {author} {\bibfnamefont
  {L.}~\bibnamefont {Li}}, \bibinfo {author} {\bibfnamefont {N.-L.}\
  \bibnamefont {Liu}}, \bibinfo {author} {\bibfnamefont {C.-Y.}\ \bibnamefont
  {Lu}}, \bibinfo {author} {\bibfnamefont {A.}~\bibnamefont {Zeilinger}}, \
  and\ \bibinfo {author} {\bibfnamefont {J.-W.}\ \bibnamefont {Pan}},\ }\href
  {\doibase 10.1103/PhysRevLett.123.070505} {\bibfield  {journal} {\bibinfo
  {journal} {Phys. Rev. Lett.}\ }\textbf {\bibinfo {volume} {123}},\ \bibinfo
  {pages} {070505} (\bibinfo {year} {2019})}\BibitemShut {NoStop}%
\bibitem [{\citenamefont {Zhang}\ \emph {et~al.}(2019)\citenamefont {Zhang},
  \citenamefont {Chen}, \citenamefont {Cui}, \citenamefont {Dowling},
  \citenamefont {Ou},\ and\ \citenamefont {Byrnes}}]{zhang:etal:pra19}%
  \BibitemOpen
  \bibfield  {author} {\bibinfo {author} {\bibfnamefont {C.}~\bibnamefont
  {Zhang}}, \bibinfo {author} {\bibfnamefont {J.~F.}\ \bibnamefont {Chen}},
  \bibinfo {author} {\bibfnamefont {C.}~\bibnamefont {Cui}}, \bibinfo {author}
  {\bibfnamefont {J.~P.}\ \bibnamefont {Dowling}}, \bibinfo {author}
  {\bibfnamefont {Z.~Y.}\ \bibnamefont {Ou}}, \ and\ \bibinfo {author}
  {\bibfnamefont {T.}~\bibnamefont {Byrnes}},\ }\href {\doibase
  10.1103/PhysRevA.100.032330} {\bibfield  {journal} {\bibinfo  {journal}
  {Phys. Rev. A}\ }\textbf {\bibinfo {volume} {100}},\ \bibinfo {pages}
  {032330} (\bibinfo {year} {2019})}\BibitemShut {NoStop}%
\bibitem [{\citenamefont {Erhard}\ \emph {et~al.}(2020)\citenamefont {Erhard},
  \citenamefont {Krenn},\ and\ \citenamefont
  {Zeilinger}}]{erhard:krenn:zeilinger:nrp20}%
  \BibitemOpen
  \bibfield  {author} {\bibinfo {author} {\bibfnamefont {M.}~\bibnamefont
  {Erhard}}, \bibinfo {author} {\bibfnamefont {M.}~\bibnamefont {Krenn}}, \
  and\ \bibinfo {author} {\bibfnamefont {A.}~\bibnamefont {Zeilinger}},\ }\href
  {\doibase 10.1038/s42254-020-0193-5} {\bibfield  {journal} {\bibinfo
  {journal} {Nature Reviews Physics}\ }\textbf {\bibinfo {volume} {2}},\
  \bibinfo {pages} {365} (\bibinfo {year} {2020})}\BibitemShut {NoStop}%
\bibitem [{\citenamefont {Hu}\ \emph {et~al.}(2020)\citenamefont {Hu},
  \citenamefont {Zhang}, \citenamefont {Liu}, \citenamefont {Cai},
  \citenamefont {Ye}, \citenamefont {Guo}, \citenamefont {Xing}, \citenamefont
  {Huang}, \citenamefont {Huang}, \citenamefont {Li},\ and\ \citenamefont
  {Guo}}]{hu:etal:prl20}%
  \BibitemOpen
  \bibfield  {author} {\bibinfo {author} {\bibfnamefont {X.-M.}\ \bibnamefont
  {Hu}}, \bibinfo {author} {\bibfnamefont {C.}~\bibnamefont {Zhang}}, \bibinfo
  {author} {\bibfnamefont {B.-H.}\ \bibnamefont {Liu}}, \bibinfo {author}
  {\bibfnamefont {Y.}~\bibnamefont {Cai}}, \bibinfo {author} {\bibfnamefont
  {X.-J.}\ \bibnamefont {Ye}}, \bibinfo {author} {\bibfnamefont
  {Y.}~\bibnamefont {Guo}}, \bibinfo {author} {\bibfnamefont {W.-B.}\
  \bibnamefont {Xing}}, \bibinfo {author} {\bibfnamefont {C.-X.}\ \bibnamefont
  {Huang}}, \bibinfo {author} {\bibfnamefont {Y.-F.}\ \bibnamefont {Huang}},
  \bibinfo {author} {\bibfnamefont {C.-F.}\ \bibnamefont {Li}}, \ and\ \bibinfo
  {author} {\bibfnamefont {G.-C.}\ \bibnamefont {Guo}},\ }\href {\doibase
  10.1103/PhysRevLett.125.230501} {\bibfield  {journal} {\bibinfo  {journal}
  {Phys. Rev. Lett.}\ }\textbf {\bibinfo {volume} {125}},\ \bibinfo {pages}
  {230501} (\bibinfo {year} {2020})}\BibitemShut {NoStop}%
\bibitem [{\citenamefont {Ren}\ \emph {et~al.}(2017)\citenamefont {Ren} \emph
  {et~al.}}]{ren:etal:nat17}%
  \BibitemOpen
  \bibfield  {author} {\bibinfo {author} {\bibfnamefont {J.-G.}\ \bibnamefont
  {Ren}} \emph {et~al.},\ }\href {\doibase 10.1038/nature23675} {\bibfield
  {journal} {\bibinfo  {journal} {Nature}\ }\textbf {\bibinfo {volume} {549}},\
  \bibinfo {pages} {70} (\bibinfo {year} {2017})}\BibitemShut {NoStop}%
\bibitem [{\citenamefont {Bennett}\ and\ \citenamefont
  {Brassard}(2014)}]{bennett:brassard:proceed84}%
  \BibitemOpen
  \bibfield  {author} {\bibinfo {author} {\bibfnamefont {C.~H.}\ \bibnamefont
  {Bennett}}\ and\ \bibinfo {author} {\bibfnamefont {G.}~\bibnamefont
  {Brassard}}\ }(\bibinfo  {publisher} {Elsevier BV},\ \bibinfo {year} {2014})\
  p.\ \bibinfo {pages} {7–11}\BibitemShut {NoStop}%
\bibitem [{\citenamefont {Ekert}(1991)}]{ekert:prl91}%
  \BibitemOpen
  \bibfield  {author} {\bibinfo {author} {\bibfnamefont {A.~K.}\ \bibnamefont
  {Ekert}},\ }\href {\doibase 10.1103/PhysRevLett.67.661} {\bibfield  {journal}
  {\bibinfo  {journal} {Phys. Rev. Lett.}\ }\textbf {\bibinfo {volume} {67}},\
  \bibinfo {pages} {661} (\bibinfo {year} {1991})}\BibitemShut {NoStop}%
\bibitem [{\citenamefont {Cirac}\ \emph {et~al.}(1997)\citenamefont {Cirac},
  \citenamefont {Zoller}, \citenamefont {Kimble},\ and\ \citenamefont
  {Mabuchi}}]{cirac:zoller:kimble:mabuchi:prl97}%
  \BibitemOpen
  \bibfield  {author} {\bibinfo {author} {\bibfnamefont {J.~I.}\ \bibnamefont
  {Cirac}}, \bibinfo {author} {\bibfnamefont {P.}~\bibnamefont {Zoller}},
  \bibinfo {author} {\bibfnamefont {H.~J.}\ \bibnamefont {Kimble}}, \ and\
  \bibinfo {author} {\bibfnamefont {H.}~\bibnamefont {Mabuchi}},\ }\href
  {\doibase 10.1103/PhysRevLett.78.3221} {\bibfield  {journal} {\bibinfo
  {journal} {Phys. Rev. Lett.}\ }\textbf {\bibinfo {volume} {78}},\ \bibinfo
  {pages} {3221} (\bibinfo {year} {1997})}\BibitemShut {NoStop}%
\bibitem [{\citenamefont {Kimble}(2008)}]{kimble:nat08}%
  \BibitemOpen
  \bibfield  {author} {\bibinfo {author} {\bibfnamefont {H.~J.}\ \bibnamefont
  {Kimble}},\ }\href {\doibase 10.1038/nature07127} {\bibfield  {journal}
  {\bibinfo  {journal} {Nature}\ }\textbf {\bibinfo {volume} {453}},\ \bibinfo
  {pages} {1023} (\bibinfo {year} {2008})}\BibitemShut {NoStop}%
\bibitem [{\citenamefont {Wehner}\ \emph {et~al.}(2018)\citenamefont {Wehner},
  \citenamefont {Elkouss},\ and\ \citenamefont
  {Hanson}}]{wehner:elkouss:hanson:sc18}%
  \BibitemOpen
  \bibfield  {author} {\bibinfo {author} {\bibfnamefont {S.}~\bibnamefont
  {Wehner}}, \bibinfo {author} {\bibfnamefont {D.}~\bibnamefont {Elkouss}}, \
  and\ \bibinfo {author} {\bibfnamefont {R.}~\bibnamefont {Hanson}},\ }\href
  {\doibase 10.1126/science.aam9288} {\bibfield  {journal} {\bibinfo  {journal}
  {Science}\ }\textbf {\bibinfo {volume} {362}},\ \bibinfo {pages} {eaam928}
  (\bibinfo {year} {2018})}\BibitemShut {NoStop}%
\bibitem [{\citenamefont {Raussendorf}\ and\ \citenamefont
  {Briegel}(2001)}]{raussendorf:briegel:prl01}%
  \BibitemOpen
  \bibfield  {author} {\bibinfo {author} {\bibfnamefont {R.}~\bibnamefont
  {Raussendorf}}\ and\ \bibinfo {author} {\bibfnamefont {H.~J.}\ \bibnamefont
  {Briegel}},\ }\href {\doibase 10.1103/PhysRevLett.86.5188} {\bibfield
  {journal} {\bibinfo  {journal} {Phys. Rev. Lett.}\ }\textbf {\bibinfo
  {volume} {86}},\ \bibinfo {pages} {5188} (\bibinfo {year}
  {2001})}\BibitemShut {NoStop}%
\bibitem [{\citenamefont {Briegel}\ \emph {et~al.}(2009)\citenamefont
  {Briegel}, \citenamefont {Browne}, \citenamefont {D{\"u}r}, \citenamefont
  {Raussendorf},\ and\ \citenamefont {Van~den
  Nest}}]{briegel:raussendorf:nest:nphys09}%
  \BibitemOpen
  \bibfield  {author} {\bibinfo {author} {\bibfnamefont {H.~J.}\ \bibnamefont
  {Briegel}}, \bibinfo {author} {\bibfnamefont {D.~E.}\ \bibnamefont {Browne}},
  \bibinfo {author} {\bibfnamefont {W.}~\bibnamefont {D{\"u}r}}, \bibinfo
  {author} {\bibfnamefont {R.}~\bibnamefont {Raussendorf}}, \ and\ \bibinfo
  {author} {\bibfnamefont {M.}~\bibnamefont {Van~den Nest}},\ }\href {\doibase
  10.1038/nphys1157} {\bibfield  {journal} {\bibinfo  {journal} {Nature
  Physics}\ }\textbf {\bibinfo {volume} {5}},\ \bibinfo {pages} {19} (\bibinfo
  {year} {2009})}\BibitemShut {NoStop}%
\bibitem [{\citenamefont {Briegel}\ \emph {et~al.}(1998)\citenamefont
  {Briegel}, \citenamefont {D\"ur}, \citenamefont {Cirac},\ and\ \citenamefont
  {Zoller}}]{briegel:dur:cirac:zoller:prl98}%
  \BibitemOpen
  \bibfield  {author} {\bibinfo {author} {\bibfnamefont {H.-J.}\ \bibnamefont
  {Briegel}}, \bibinfo {author} {\bibfnamefont {W.}~\bibnamefont {D\"ur}},
  \bibinfo {author} {\bibfnamefont {J.~I.}\ \bibnamefont {Cirac}}, \ and\
  \bibinfo {author} {\bibfnamefont {P.}~\bibnamefont {Zoller}},\ }\href
  {\doibase 10.1103/PhysRevLett.81.5932} {\bibfield  {journal} {\bibinfo
  {journal} {Phys. Rev. Lett.}\ }\textbf {\bibinfo {volume} {81}},\ \bibinfo
  {pages} {5932} (\bibinfo {year} {1998})}\BibitemShut {NoStop}%
\bibitem [{\citenamefont {Nielsen}\ and\ \citenamefont
  {Chuang}(2010)}]{book:mike:ike:cambridge10}%
  \BibitemOpen
  \bibfield  {author} {\bibinfo {author} {\bibfnamefont {M.~A.}\ \bibnamefont
  {Nielsen}}\ and\ \bibinfo {author} {\bibfnamefont {I.~L.}\ \bibnamefont
  {Chuang}},\ }\href {https://doi.org/10.1017/CBO9780511976667} {\emph
  {\bibinfo {title} {Quantum Computation and Quantum Information: 10th
  Anniversary Edition}}}\ (\bibinfo  {publisher} {Cambridge University Press},\
  \bibinfo {address} {Cambridge},\ \bibinfo {year} {2010})\BibitemShut
  {NoStop}%
\bibitem [{\citenamefont {Wootters}\ and\ \citenamefont
  {Zurek}(1982)}]{wootters:zurek:nat82}%
  \BibitemOpen
  \bibfield  {author} {\bibinfo {author} {\bibfnamefont {W.~K.}\ \bibnamefont
  {Wootters}}\ and\ \bibinfo {author} {\bibfnamefont {W.~H.}\ \bibnamefont
  {Zurek}},\ }\href {\doibase 10.1038/299802a0} {\bibfield  {journal} {\bibinfo
   {journal} {Nature}\ }\textbf {\bibinfo {volume} {299}},\ \bibinfo {pages}
  {802} (\bibinfo {year} {1982})}\BibitemShut {NoStop}%
\bibitem [{\citenamefont {Nunavath}\ \emph {et~al.}(2024)\citenamefont
  {Nunavath}, \citenamefont {Mishra},\ and\ \citenamefont
  {Pathak}}]{nunavath:mishra:pathak:arxiv2408}%
  \BibitemOpen
  \bibfield  {author} {\bibinfo {author} {\bibfnamefont {N.}~\bibnamefont
  {Nunavath}}, \bibinfo {author} {\bibfnamefont {S.}~\bibnamefont {Mishra}}, \
  and\ \bibinfo {author} {\bibfnamefont {A.}~\bibnamefont {Pathak}},\ }\href
  {https://arxiv.org/abs/2408.16016} {\enquote {\bibinfo {title} {Quantum
  teleportation using quantum candies},}\ } (\bibinfo {year} {2024}),\ \Eprint
  {http://arxiv.org/abs/2408.16016} {arXiv:2408.16016 [physics.pop-ph]}
  \BibitemShut {NoStop}%
\bibitem [{\citenamefont {Lin}\ and\ \citenamefont
  {Mor}(2020)}]{lin:mor:proceed20}%
  \BibitemOpen
  \bibfield  {author} {\bibinfo {author} {\bibfnamefont {J.}~\bibnamefont
  {Lin}}\ and\ \bibinfo {author} {\bibfnamefont {T.}~\bibnamefont {Mor}},\ }in\
  \href {https://link.springer.com/chapter/10.1007/978-3-030-63000-3_6} {\emph
  {\bibinfo {booktitle} {Theory and Practice of Natural Computing}}}\ (\bibinfo
   {publisher} {Springer International Publishing},\ \bibinfo {address}
  {Cham},\ \bibinfo {year} {2020})\ pp.\ \bibinfo {pages} {69--81}\BibitemShut
  {NoStop}%
\bibitem [{\citenamefont {Lin}\ \emph {et~al.}(2021)\citenamefont {Lin},
  \citenamefont {Mor},\ and\ \citenamefont
  {Shapira}}]{lin:mor:shapira:arxiv2110}%
  \BibitemOpen
  \bibfield  {author} {\bibinfo {author} {\bibfnamefont {J.}~\bibnamefont
  {Lin}}, \bibinfo {author} {\bibfnamefont {T.}~\bibnamefont {Mor}}, \ and\
  \bibinfo {author} {\bibfnamefont {R.}~\bibnamefont {Shapira}},\ }\href
  {https://arxiv.org/abs/2110.01402} {\enquote {\bibinfo {title} {Quantum
  information and beyond -- with quantum candies},}\ } (\bibinfo {year}
  {2021}),\ \Eprint {http://arxiv.org/abs/2110.01402} {arXiv:2110.01402
  [physics.ed-ph]} \BibitemShut {NoStop}%
\bibitem [{\citenamefont {Bennett}\ and\ \citenamefont
  {Wiesner}(1992)}]{bennet:wiesner:prl92}%
  \BibitemOpen
  \bibfield  {author} {\bibinfo {author} {\bibfnamefont {C.~H.}\ \bibnamefont
  {Bennett}}\ and\ \bibinfo {author} {\bibfnamefont {S.~J.}\ \bibnamefont
  {Wiesner}},\ }\href {\doibase 10.1103/PhysRevLett.69.2881} {\bibfield
  {journal} {\bibinfo  {journal} {Phys. Rev. Lett.}\ }\textbf {\bibinfo
  {volume} {69}},\ \bibinfo {pages} {2881} (\bibinfo {year}
  {1992})}\BibitemShut {NoStop}%
\bibitem [{\citenamefont {\ifmmode~\dot{Z}\else \.{Z}\fi{}ukowski}\ \emph
  {et~al.}(1993)\citenamefont {\ifmmode~\dot{Z}\else \.{Z}\fi{}ukowski},
  \citenamefont {Zeilinger}, \citenamefont {Horne},\ and\ \citenamefont
  {Ekert}}]{zukowski:zeilinger:horne:ekert:prl93}%
  \BibitemOpen
  \bibfield  {author} {\bibinfo {author} {\bibfnamefont {M.}~\bibnamefont
  {\ifmmode~\dot{Z}\else \.{Z}\fi{}ukowski}}, \bibinfo {author} {\bibfnamefont
  {A.}~\bibnamefont {Zeilinger}}, \bibinfo {author} {\bibfnamefont {M.~A.}\
  \bibnamefont {Horne}}, \ and\ \bibinfo {author} {\bibfnamefont {A.~K.}\
  \bibnamefont {Ekert}},\ }\href {\doibase 10.1103/PhysRevLett.71.4287}
  {\bibfield  {journal} {\bibinfo  {journal} {Phys. Rev. Lett.}\ }\textbf
  {\bibinfo {volume} {71}},\ \bibinfo {pages} {4287} (\bibinfo {year}
  {1993})}\BibitemShut {NoStop}%
\end{thebibliography}%


\end{document}